\documentclass[aps,prd,showpacs,twocolumn]{revtex4}
\usepackage{graphicx}

\newcommand{\ga}{\stackrel{>}{ _{\sim}}}
\newcommand{\ve}[1]{\mbox{\boldmath $#1$}}
\newcommand{\beq}{\begin{equation}}
\newcommand{\eeq}{\end{equation}}
\newcommand{\beqn}{\begin{eqnarray}}
\newcommand{\eeqn}{\end{eqnarray}}
\newcommand{\mcal}[1]{{\mathcal #1}}
\newcommand{\bit}{\begin{itemize}}

\begin{document}
\title{Magnetic Braking in Differentially Rotating, Relativistic Stars}

\author{Yuk Tung Liu}

\author{Stuart L.\ Shapiro}
\altaffiliation{Department of Astronomy \& NCSA, University of Illinois 
at Urbana-Champaign, Urbana, IL 61801}

\affiliation{Department of Physics, University of Illinois at
Urbana-Champaign, Urbana, IL~61801}

\begin{abstract}
We study the magnetic braking and viscous damping of differential rotation in
incompressible,
uniform density stars in general relativity. Differentially rotating stars can
support
significantly more mass in equilibrium than nonrotating or uniformly rotating
stars,
according to general relativity. The remnant of a binary neutron star merger or
supernova core
collapse may produce such a ``hypermassive'' neutron star. Although 
a hypermassive neutron star may be stable on a
dynamical timescale, magnetic braking and viscous damping of differential
rotation will
ultimately alter the equilibrium structure, possibly leading to delayed
catastrophic collapse.
Here we treat the slow-rotation, weak-magnetic field limit
in which $E_{\rm rot} \ll E_{\rm mag} \ll W$, where $E_{\rm rot}$ is the
rotational
kinetic energy, $E_{\rm mag}$ is the magnetic energy, and $W$ is the
gravitational binding energy of the star. 
We assume the system to be axisymmetric and solve the MHD
equations in both Newtonian gravitation and general relativity. For an
initially uniform
magnetic field parallel to the rotation axis in which we
neglect viscosity, the Newtonian case can be solved analytically, but the other
cases we consider require a numerical integration.
Toroidal magnetic fields are generated whenever the angular velocity varies
along the initial poloidal field lines. We find that
the toroidal fields and angular velocities oscillate independently along each
poloidal field line, which enables us to transform the original 2+1 equations
into 1+1 form and solve them along each field line independently.
The incoherent oscillations on different field lines
stir up turbulent-like motion in tens of Alfv\'en timescales (``phase mixing''). 
In the presence of viscosity, the stars eventually are
driven
to uniform rotation, with the energy contained in the initial differential
rotation
going into heat. Our evolution calculations serve as qualitative guides and
benchmarks for future, more realistic MHD simulations in full 3+1 general
relativity.
\end{abstract}


\maketitle

\section{Introduction}

New-born neutron stars formed from core collapse supernovae, accretion 
induced collapse of white dwarfs, and coalescence of neutron star binaries
are likely to be differentially 
rotating~\cite{zwerger97,ruffert99,liu01,liu02,rasio929499,shibata00}. 
Differential rotation will cause a frozen-in poloidal magnetic field to wind up 
and generate a strong toroidal field. The back-reaction of the magnetic stresses 
on the fluid motion -- magnetic braking -- will 
destroy the initial differential rotation and can result in a significant 
change in the structure and dynamics of the star.

Differentially rotating neutron stars can support significantly more rest mass
than their nonrotating or uniformly rotating counterparts, creating ``hypermassive'' 
neutron stars~\cite{baumgarte00,lyford03}. Such hypermassive neutron stars can 
form from the coalescence of neutron star binaries~\cite{rasio929499,shibata00}. 
The stabilization arising from differential rotation, although 
expected to last for many dynamical timescales, will ultimately be destroyed 
by magnetic braking and/or viscosity~\cite{baumgarte00,shapiro00}. These processes 
drive the star to 
uniform rotation, which cannot support the excess mass, and can lead to ``delayed''
catastrophic collapse, possibly accompanied by some mass loss.

Conservation of angular momentum implies that the nascent neutron stars 
formed from core collapse supernovae or accretion induced collapse of 
white dwarfs are rapidly and differentially rotating. These neutron stars 
may develop a nonaxisymmetric bar-mode instability that might 
be detectable as a quasi-periodic gravitational wave by gravitational wave 
detectors, such as LIGO, VIRGO, GEO and TAMA. 
The {\em dynamical} bar-instability usually develops when the ratio $\beta=T/W$ 
is high enough, e.g.\ $\beta \ga$ 0.25 -- 0.27~\cite{rampp98,liu02}, 
where $T$ is the rotational kinetic energy and W is the gravitational 
binding energy (however, 
for extreme degrees of differential 
rotation a bar-instability may also develop for stars with very low 
$\beta$~\cite{shibata02,shibata03,karino03}; moreover, an $m=1$ 
one-armed spiral mode may also become dynamically unstable for stars with a very 
soft equation of state and a high degree of differential 
rotation~\cite{centrella01,saijo03}). 
A hot, proto-neutron star formed a few milliseconds 
after core bounce may not have a sufficiently high value of 
$\beta$ to trigger the dynamical bar-instability immediately. 
However, the instability might develop after 
about 20 seconds, after which neutrinos will have carried away 
most of the thermal energy, causing the star to contract further 
and $\beta$ to exceed the threshold value 
for the bar-instability. If the proto-neutron star has a strong 
magnetic field ($B \ga 10^{12} G$), magnetic braking could significantly 
change the angular momentum distribution during the neutrino cooling phase 
and might suppress the bar-instability~\cite{liu02}. On the other hand, 
the {\em secular} bar-instability could develop at much lower 
$\beta$~\cite{sec-bar-notes} on a longer timescale. 
Once again, a small, frozen-in seed magnetic field could be wound up 
to sufficiently high strength by differential rotation to suppress 
the secular instability over this timescale.

Short-duration gamma-ray bursts (GRBs) are thought to result from binary 
neutron star mergers~\cite{narayan92}, or tidal disruptions of neutron stars
by black holes~\cite{ruffert99}. Long-duration GRBs likely result from 
the collapse of rotating, massive stars which form 
black holes, followed by supernova explosions~\cite{hjorth03,macfadyen99}. 
In current scenarios, the burst is 
powered by the extraction of rotational energy from the neutron star 
or black hole, or from the remnant disk material formed around the black 
hole~\cite{vlahakis01}. Strong magnetic fields provide the likely 
mechanism for extracting this energy on the required timescale and 
driving collimated GRB outflows in the form of relativistic 
jets~\cite{meszaros97,sari99,piran03}. Even if the initial magnetic 
fields are weak, they can be amplified to the required values by 
differential motions or dynamo action.

The $r$-mode instability has recently been proposed as a possible mechanism
for limiting the angular velocities in neutron stars and producing
observable quasi-periodic gravitational
waves~\cite{andersson98,friedman98,andersson99,lindblom98}. However,
preliminary calculations (see
Refs.~\cite{rezzolla00,rezzolla01a,rezzolla01b} and references therein)
suggest that if the stellar magnetic field is strong enough, $r$-mode
oscillations will not occur. Even if the initial field is weak, fluid
motions produced by these oscillations
may amplify the magnetic field and eventually
distort or suppress the $r$-modes altogether ($r$-mode oscillations may 
also be suppressed
by nonlinear mode coupling~\cite{arras02,schenk02,gressman02} or
hyperon bulk viscosity~\cite{jones01a,jones01b,lindblom02}).

In a different context, supermassive stars (SMSs) may form in the 
early universe, and their 
catastrophic collapse may provide the seeds of supermassive black 
holes (SMBHs) observed in galaxies and quasars (see 
Refs.~\cite{rees84,baumgarte99a,shapiro03} 
for discussion and references). If an SMS maintains
uniform rotation as it cools and contracts, it will ultimately 
arrive at the onset of a relativistic radial instability, 
triggering coherent dynamical collapse to an SMBH and giving rise to 
a burst of gravitational waves~\cite{saijo02,shibata02b}. If an SMS 
is differentially rotating, cooling and contraction will instead 
lead to the unstable formation of bars or spiral arms prior to 
collapse and the production of quasi-periodic gravitational 
waves~\cite{new01a,new01b}. Since
magnetic fields and turbulent viscosity provide the principal mechanisms 
that can damp differential rotation in such stars~\cite{zeldovich71,shapiro00}, 
their role is therefore crucial in determining the fate of these objects.

Motivated by the importance of magnetic fields in differentially 
rotating relativistic stars, Shapiro performed a simple, 
Newtonian, MHD calculation showing some of the effects of magnetic 
braking (\cite{shapiro00}, hereafter Paper~I). 
In his simplified model, the star is idealized 
as a differentially rotating, infinite cylinder consisting of a homogeneous, 
incompressible conducting gas. The magnetic field is taken to be purely 
radial initially and is allowed to evolve according to the ideal MHD 
equations (flux-freezing).
The calculation shows that differential rotation 
generates a toroidal magnetic field, which reacts
back on the fluid flow. In the absence of viscous dissipation, 
the toroidal field energy and rotational kinetic energy in differential 
motion undergo periodic oscillations on the Alfv\'en timescale. 
The magnitude of the oscillations is independent of the initial 
magnetic field strength; only the growth and oscillation timescale 
depends on the magnitude 
of the seed field. If viscosity is present, or if some 
of the Alfv\'en waves are allowed to propagate out of the star and into 
an ambient plasma atmosphere, the oscillations are damped, rotational energy 
is dissipated, and the star is driven to uniform rotation. 

Recently, Cook, Shapiro, and Stephens (\cite{cook03}, hereafter Paper~II) 
generalized Shapiro's 
calculations for compressible stars. In their model, the star is idealized 
as a differentially 
rotating, infinite cylinder supported by a polytropic equation of 
state. They performed Newtonian MHD simulations for differentially rotating 
stars with various polytropic indices and different values of $\beta$. 
They found that when $\beta$ is below the upper (mass-shedding) limit 
for uniform rotation, $\beta_{\rm max}$, magnetic braking results in oscillations 
of the induced toroidal fields and angular velocities, and the star pulsates
stably. However, when $\beta$ exceeds $\beta_{\rm max}$, 
their calculations suggest that the core contracts significantly, but 
quasistatically, while
the outer layers are ejected to large radii to form a wind or an ambient disk.

In this paper, we consider another idealized, but useful, model to explore the 
effects of general relativity on magnetic braking and viscous damping of 
differential rotation in stars. Our star is 
idealized as an incompressible, slowly differentially rotating 
sphere of uniform density, threaded by a seed poloidal magnetic field 
at time $t=0$. To simplify the calculations, we restrict our analysis to 
the case in which $E_{\rm rot} \ll E_{\rm mag} \ll W$, where $E_{\rm rot}$=T 
is the rotational kinetic energy, and $E_{\rm mag}$ is the magnetic energy. 
The condition $E_{\rm mag} \gg E_{\rm rot}$ is equivalent to the condition 
that the Alfv\'en timescale $t_A$ is 
much shorter than the rotation period $P_{\rm rot}$. 
The condition that $E_{\rm mag} \ll W$ guarantees that the magnetic 
field is small in comparison to the internal pressure forces and 
gravitational field.
We perform MHD evolution calculations in Newtonian gravity, first as a 
``warm-up'', and then in full general relativity, for various initial 
field configurations. For an initial magnetic field parallel to the 
rotation axis in which we neglect viscosity, the Newtonian equations 
can be solved analytically, but the other cases we consider require 
numerical integration. Adopting axisymmetry, 
our MHD equations are two-dimensional, as 
opposed to one-dimensional in Papers~I and~II. 
We found that this extra degree of freedom changes the description of magnetic 
braking. 
We found that the angular velocity and toroidal magnetic field undergo 
periodic oscillations along the initial poloidal field lines,
in the absence of viscosity. However, the oscillation 
frequencies are different along each poloidal field line. The incoherent 
oscillations on different field lines eventually destroys the laminar 
flow and creates irregular velocity fields across the poloidal 
field lines. 
This effect has been studied in Newtonian MHD and is sometimes referred to as
``phase mixing'' (see~\cite{spruit99} and references therein). 
We demonstrate that this phase mixing effect is also present in relativistic 
MHD.

In the situation where the Alfv\'en timescale is much shorter than the rotation 
period, which is the limit we consider in this paper, the phase mixing 
may create turbulent-like flows in tens of Alfv\'en timescales. 
We also 
studied the effect of magnetic braking in the presence of viscosity. We found 
that, not surprisingly, the star eventually will be driven to uniform 
rotation by the combined action of the magnetic fields and viscosity.

Although our analysis in this paper is restricted to the early phases of 
evolution in the slow-rotation, weak-magnetic field limit, we are able to 
solve the full nonlinear MHD equations for a highly relativistic, 
differentially rotating star. Our calculations serve as qualitative guides and
benchmarks for future, more realistic MHD simulations in full 3+1 general
relativity.

We structure this paper as follows: in Section~\ref{sec:model}, we 
describe our model in detail. We derive and analyze the Newtonian 
MHD equations in Section~\ref{sec:Neweq}. In Section~\ref{sec:Rel}, 
we derive the analogous relativistic MHD equations using the 3+1 formalism 
presented by Baumgarte and Shapiro~\cite{baumgarte03}. We then rewrite 
the MHD equations in nondimensional form in Section~\ref{sec:nondim}, 
and present our numerical results in Section~\ref{sec:numerics}. 
Finally, we summarize and discuss our results in Section~\ref{sec:diss}.

\section{Model}
\label{sec:model}

We consider an incompressible, rotating equilibrium star of uniform 
density (i.e.\ a polytrope of index $n=0$). At time $t=0$, the star is 
assumed to rotate slowly and differentially with a small axisymmetric 
poloidal magnetic field. 
We assume that both the rotational kinetic energy and the magnetic energy are 
small compared to the gravitational binding energy. In Newtonian gravity, 
the deviation from the spherical solution 
is of second order in the magnetic field strength and/or in the magnitude of angular 
velocity and is therefore neglected in our treatment. In general relativity, 
the leading order correction to the metric comes from the dragging of 
inertial frames, which is of first order. We therefore neglect the 
deformation of the star in Newtonian gravity, but keep the frame dragging 
term in general relativity.

As discussed in Papers~I and~II, differential rotation 
twists up the frozen-in poloidal magnetic field and generates a toroidal field 
in the star. The toroidal field causes a shear stress on the fluid and 
this stress changes the angular 
velocity profile on the Alfv\'en timescale of the poloidal field. 
Unlike the cases studied in Papers~I and~II, 
the changes in angular velocity will generate typically a meridional current. The 
interaction of magnetic field and the meridional current may develop MHD
instabilities and result in turbulence~\cite{balbus98}. 

In this paper, we consider the case in which
the Alfv\'en timescale is much shorter than the rotation period. In particular,
we consider the case in which $W \gg E_{\rm mag} \gg E_{\rm rot}$. 
The condition $E_{\rm mag} \gg E_{\rm rot}$ 
is equivalent to the condition that $v_A \gg \Omega R$, where $v_A$ is a 
typical Alfv\'en speed, $\Omega$ is a typical angular velocity inside 
the star, and $R$ is the radius of the star. 
It can be shown (see
Appendix~\ref{app:v}) that the meridional current can be ignored in the early phase
of the magnetic braking. In this paper, 
we focus on the effect of magnetic braking before the meridional current 
builds up. We assume the system remains axisymmetric on the timescale of interest. 

\section{Newtonian equations}
\label{sec:Neweq}

\subsection{Basic Equations}

We start with the MHD equations for a perfectly conducting, incompressible, 
viscous Newtonian fluid (see, e.g.~\cite{jackson75,landau87})
\beqn
  \frac{\partial \ve{B}}{\partial t} &=& \ve{\nabla} \times (\ve{v} \times 
\ve{B}) \ , \label{eq:beq1} \\ 
  \rho \left(\frac{\partial \ve{v}}{\partial t} + \ve{v} \cdot 
\ve{\nabla} \ve{v} \right) &=& -\ve{\nabla} p -\rho \ve{\nabla} \Phi
- \ve{\nabla} \left( \frac{B^2}{8\pi}\right) +\frac{1}{4\pi} 
 (\ve{B}\cdot \ve{\nabla}) \ve{B} \cr & & \cr
 & & + \ve{\nabla} \cdot (\eta \ve{\sigma}) + \ve{\nabla}
(\zeta \ve{\nabla} \cdot \ve{v}) \label{eq:beq2} \ , \\ 
  \ve{\nabla}\cdot \ve{v} &=& 0 \label{eq:beq3} \ , \\
  \nabla^2 \Phi &=& 4\pi G \rho \label{eq:beq4} \ ,
\eeqn
where $\eta$ and $\zeta$ are the coefficients of shear and bulk viscosity, 
respectively, and in general are functions of pressure and temperature. The 
components of the shear tensor in Cartesian coordinates are given 
by
\beq
   \sigma_{ij}=\partial_i v_j + \partial_j v_i -\frac{2}{3} \delta_{ij}
\ve{\nabla} \cdot \ve{v} \ .
\eeq
For an incompressible fluid, Eq.~(\ref{eq:beq3}) allows us to write
\beqn
  \frac{\partial \ve{B}}{\partial t} &=& \ve{B} \cdot \ve{\nabla} \ve{v}
  - \ve{v} \cdot \ve{\nabla} \ve{B} \ , \label{eq:B0} \\
  \frac{\partial \ve{v}}{\partial t} + \ve{v} \cdot 
\ve{\nabla} \ve{v} &=& -\frac{1}{\rho}\ve{\nabla} p - \ve{\nabla}\Phi - \ve{\nabla} 
\left(\frac{B^2}{8\pi}\right)
+\frac{1}{4\pi\rho} \ve{B} \cdot \ve{\nabla} \ve{B} \cr  & & \cr
 & & + \ve{\nabla} \cdot (\nu \ve{\sigma}) \ ,
\label{eq:MHD0}
\eeqn
where $\nu=\eta/\rho$. 

In order to make a direct comparison between Newtonian and relativistic MHD 
equations, we adopt some conventions used in relativity. 
For the spherical coordinate system $(r,\theta,\phi)$, we introduce 
three orthonormal unit basis vectors 
\beqn
  \ve{e}_{\hat{r}} &=& \sin \theta \cos \phi \ve{e}_x
 + \sin \theta \sin \phi \ve{e}_y + \cos \theta \ve{e}_z \\
  \ve{e}_{\hat{\theta}} &=& \cos \theta \cos \phi \ve{e}_x  
+ \cos \theta \sin \phi \ve{e}_y - \sin \theta \ve{e}_z \\
  \ve{e}_{\hat{\phi}} &=& -\sin \phi \ve{e}_x + \cos \phi \ve{e}_y  \ ,
\eeqn
where $\ve{e}_x$, $\ve{e}_y$ and $\ve{e}_z$ are the usual Cartesian 
unit basis vectors. The basis vectors satisfy $\ve{e}_{\hat{i}} \cdot 
\ve{e}_{\hat{j}} =\delta_{ij}$, where $i$ and $j$ denote $r$, $\theta$ 
and $\phi$. Any vector $\ve{V}$ can be expanded in these three basis vectors as 
\beq
  \ve{V} = V^{\hat{r}} \ve{e}_{\hat{r}} + V^{\hat{\theta}} 
\ve{e}_{\hat{\theta}} + V^{\hat{\phi}} \ve{e}_{\hat{\phi}} \ .
\eeq
We also define the coordinate basis vectors 
\beq
  \ve{e}_r=\ve{e}_{\hat{r}} \ , \ \ \ve{e}_{\theta}=r\ve{e}_{\hat{\theta}} 
\ , \ \ \ve{e}_{\phi}=r\sin \theta \ve{e}_{\hat{\phi}} \ ,
\eeq
which satisfy $\ve{e}_i \cdot \ve{e}_j = g_{ij}$. Here the spatial metric tensor 
satisfies
\beq
  g_{rr} = 1 \ , \ \ g_{\theta \theta}=r^2 \ , \ \ g_{\phi \phi}=r^2 \sin^2 
\theta \ ,
\eeq
and all the off-diagonal components of $g_{ij}$ are zeros. Any vector $\ve{V}$ 
can be expanded in these coordinate basis vectors as
\beq
  \ve{V}= V^r \ve{e}_r + V^{\theta} \ve{e}_{\theta} + V^{\phi} \ve{e}_{\phi} \ ,
\eeq
hence we have
\beq
  V^r = V^{\hat{r}} \ , \ \ V^{\theta} = V^{\hat{\theta}}/r \ , \ \ 
  V^{\phi} = V^{\hat{\phi}}/r \sin \theta \ .
\eeq

We assume the system 
is axisymmetric and ignore the meridional components of velocity 
(see Appendix~\ref{app:v}). We therefore
set $v^r=v^{\theta}=0$, and assume there is no toroidal field at time $t=0$, 
$B^{\phi}(0;r,\theta)=0$. We also assume that the star has a small amount of 
rotation $v^{\phi}=\Omega(0;r,\theta)$ and an initial poloidal field 
$\ve{B}(0;r,\theta)=B^r(0;r,\theta) \ve{e_r} + B^{\theta}(0;r,\theta) 
\ve{e_{\theta}}$. Equation~(\ref{eq:B0}) immediately 
gives $\partial_t B^r=\partial_t B^{\theta}=0$. Hence the poloidal 
field, which we shall designate $\ve{B}(0)$, does not evolve with time. 
Equations~(\ref{eq:B0}) and~(\ref{eq:MHD0}) simplify to yield
\beqn
   \partial_t B^{\phi} &=& B^j(0) \partial_j \Omega  \ , \label{eq:B1} \\
   \partial_t \Omega &=& \frac{1}{4 \pi \rho r^2\sin^2 \theta} B^j(0) 
\partial_j (r^2 \sin^2 \theta B^{\phi}) + (\partial_t \Omega)_{\rm vis}
\label{eq:MHD1} \ , \ \ \ \ \ \ 
\eeqn
where $j$ denotes $r$ and $\theta$, and the usual summation convention is adopted. 
In spherical coordinates, this viscosity term takes the form
\beq
  (\partial_t \Omega)_{\rm vis} = \frac{1}{r^4}
\partial_{r} (\nu r^4
\partial_{r} \Omega) + \frac{1}{r^2 \sin^3 \theta}
\partial_{\theta} ( \nu \sin^3 \theta \partial_{\theta}
\Omega) \ .
\label{eq:vis}
\eeq

The toroidal field is generated via Alfv\'en waves, which cannot 
propagate in vacuum. Hence the toroidal 
field cannot be carried outside the star. This fact, together with 
Eq.~(\ref{eq:B1}), leads to the boundary condition
\beq
  B^{\phi}(t,R,\theta) = 0 = \left. B^j(0)\partial_j \Omega \right|_{r=R} \ ,
\label{eq:BC}
\eeq
where $R$ is the radius of the star.
 
\subsection{Initial Magnetic Field}
\label{sec:IMF}

In order to solve the MHD equations~(\ref{eq:B1}) and~(\ref{eq:MHD1}), we 
need to specify the poloidal field $\ve{B}(0)$ inside the star.
In this paper, we consider simple models of the internal magnetic field. 
A systematic way to generate an axisymmetric poloidal magnetic field 
is to assume that $\ve{\nabla}\times \ve{B}(0)=0$ inside the star. 
This corresponds to assuming that no internal electromagnetic currents are present 
in the stellar interior. Such a curl-free magnetic 
field can be generated only by a toroidal current on the surface of the star. 
It follows that we can write $\ve{B}(0)=\ve{\nabla}\Phi_m$, where 
$\Phi_m$ is a scalar function. 
The constraint $\ve{\nabla} \cdot \ve{B}(0)=0$ implies $\nabla^2 \Phi_m=0$. 
The general axisymmetric solutions to this equation that are regular at 
the origin can be expanded in terms of Legendre polynomials according to
\beq
  \Phi_m(r,\theta) = \sum_{l=0}^{\infty} a_l r^l P_l(\cos \theta) \ ,
\eeq
where $a_l$ are constants. 

\begin{figure}
\includegraphics[width=8cm]{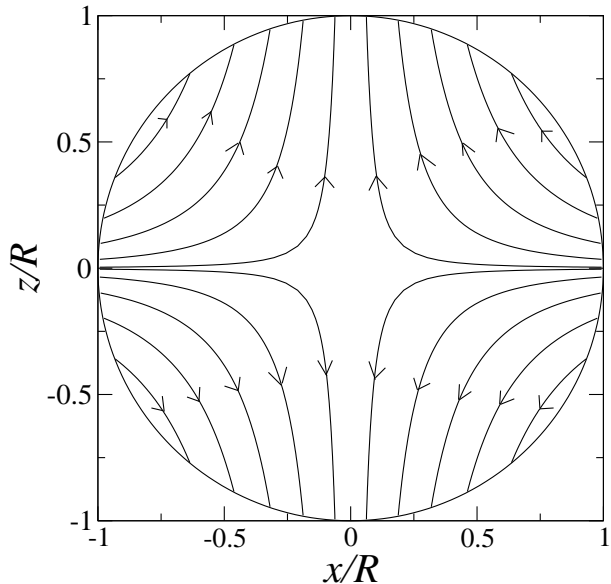}
\vskip 0.5cm
\caption{Field lines of the $l=2$ poloidal magnetic field on the $y=0$ 
(meridional) plane.}
\label{fig:L2field}
\end{figure}

The $l=0$ mode corresponds to the trivial solution 
$\ve{B}(0)=0$. For $l=1$, we have $\Phi_m = B_0 r \cos \theta$, where $B_0$ 
is a constant. Hence 
$B^{\hat{r}}(0)=B_0 \cos \theta$ and $B^{\hat{\theta}}(0)=-B_0 \sin \theta$, or 
$\ve{B}(0)=B_0 \ve{e}_z$, which is a uniform field along the rotation axis 
of the star. It follows from Eqs.~(\ref{eq:B1}) and~(\ref{eq:MHD1}) that 
if the angular velocity is constant on cylinders, i.e.\ $\partial_z \Omega=0$, 
there will be no magnetic braking in the absence of viscosity.
However, it can be proven (see, e.g.\ Ref.~\cite{tassoul78}, Section~4.3) 
that a barotropic star in rotational equilibrium must have 
such a rotation profile. 
Numerical simulations of core collapse supernovae 
also suggest that the resulting neutron stars have angular 
velocity approximately constant on cylinders (see e.g.,~\cite{janka89}). Hence we 
also consider the $l=2$ field in which $\Phi_m=B_0 r^2 (3\cos^2 \theta -1)/R$, 
where $R$ is the radius of the star. The associated magnetic field is 
\beqn
  B^{\hat{r}} &=& B_0 \left(\frac{r}{R}\right) (3\cos^2 \theta -1) \\
  B^{\hat{\theta}} &=& -3B_0 \left(\frac{r}{R}\right) \sin \theta \cos
\theta \ .
\eeqn
Figure~\ref{fig:L2field} shows the field lines for this $l=2$ field. 
Magnetic braking will occur in this case even when $\partial_z \Omega=0$. 
As will be seen later, the MHD equations for the $l=1$ field is much 
simpler than the equations for the $l=2$ field. For pedagogical purpose, 
we find it useful to study the magnetic braking for the $l=1$ 
field first, where we adopt an {\it ad hoc} rotation law with 
$\partial_z \Omega \neq 0$. We then study magnetic braking for the $l=2$ 
field with a more realistic rotation law.

\vskip 1cm

It is easy to prove that any axisymmetric poloidal 
field can be generated by a vector potential of the form 
\beq
  \ve{A}(\ve{x})= A^{\hat{\phi}}(r,\theta) \ve{e}_{\hat{\phi}} \ .
\eeq
For the $l=1$ field, we have
\beq
  A^{\hat{\phi}}(r,\theta)=\frac{1}{2} B_0 r \sin \theta \ .
\label{eq:Al1}
\eeq
For the $l=2$ field, we have 
\beq
  A^{\hat{\phi}}(r,\theta) = B_0 \left(\frac{r^2}{R}\right)\sin \theta \cos \theta \ .
\label{eq:Al2}
\eeq
The vector potential will be useful in Section~\ref{sec:sssNet}. It is also 
a convenient way to generate axisymmetric poloidal fields in relativistic 
case (see Section~\ref{sec:RelIMF}).

\subsection{Conserved Integrals}

The MHD equations~(\ref{eq:B1}) and~(\ref{eq:MHD1})
admit two non-trivial integrals of motion, one expressing the conservation of
energy and the other conservation of angular momentum.

Multiplying Eq.~(\ref{eq:MHD1}) by $\rho r^2 \sin^2 \theta$, integrating over the
volume of the star and using the boundary condition $B^{\phi}(r=R)=0$ and the
Maxwell equation $\ve{\nabla}\cdot \ve{B}(0)=0$, we find that the angular
momentum
\beq
  J = \int \rho \Omega r^2 \sin^2 \theta \, dV
\label{eq:angJ}
\eeq
is conserved.

To derive the energy integral, we first multiply Eq.~(\ref{eq:B1}) by 
$r^2 \sin^2 \theta B^{\phi}/4\pi$ and integrate over the volume of the star. 
We obtain, after integration by parts, 
\beq
  \frac{d}{dt} \int \frac{ (B^{\hat{\phi}})^2}{8\pi}\, dV = 
  -\frac{1}{4\pi} \int \Omega B^j(0) \partial_j (r^2 \sin^2 \theta B^{\phi})\, dV \ ,
\eeq
where we have used $B^{\hat{\phi}}=r\sin \theta B^{\phi}$, the boundary condition 
$B^{\phi}(r=R)=0$ and $\ve{\nabla}\cdot \ve{B}(0)=0$. Multiplying Eq.~(\ref{eq:B1}) 
by $\rho \Omega r^2 \sin^2 \theta$ and integrating over the volume of 
the star, we obtain 
\beqn
  \frac{d}{dt} \int \frac{1}{2} \rho \Omega^2 r^2 \sin^2 \theta \, dV 
  = \frac{1}{4\pi} \int B^j(0) \partial_j (r^2 \sin^2 \theta B^{\phi}) \, dV & & \cr
 + \int \rho r^2 \sin^2 \theta (\partial_t \Omega)_{\rm vis} \, dV \ . \ \ \ \ 
\ \ \ \  & & 
\eeqn
Hence we have 
\beq
  \frac{d}{dt} (E_{\rm rot} + E_{\rm mag}) = \int \rho \Omega r^2 \sin^2 \theta 
(\partial_t \Omega)_{\rm vis} \, dV \ ,
\eeq
where 
\beqn
  E_{\rm rot} &=& \int \frac{1}{2} \rho \Omega^2 r^2 \sin^2 \theta \, dV \ , \\
  E_{\rm mag} &=& \int \frac{ (B^{\hat{\phi}})^2}{8\pi}\, dV \ .
\eeqn
Note that we only include the magnetic energy associated with the toroidal 
field in $E_{\rm mag}$. One could also include the energy associated with 
the poloidal field  and still has the energy conservation since the poloidal 
field does not change with time. If we define 
\beq
  E_{\rm vis}(t) = \int_0^t \dot{E}_{\rm vis}(t') dt' \ ,
\eeq
where 
\beq
  \dot{E}_{\rm vis} = \int \rho \Omega r^2 \sin^2 \theta
(\partial_t \Omega)_{\rm vis} \, dV \ ,
\eeq
we have 
\beq
\frac{d}{dt} (E_{\rm rot} + E_{\rm mag} + E_{\rm vis}) = 0 \ .
\label{eq:Engterms}
\eeq
Hence the total energy $E=E_{\rm mag} + E_{\rm mag} + E_{\rm vis}$ is conserved. 

To compute $\dot{E}_{\rm vis}$, we use Eq.~(\ref{eq:vis}) for 
$(\partial_t \Omega)_{\rm vis}$. After integration by parts, we obtain 
\beqn
  \dot{E}_{\rm vis} &=& \int \eta r^2 \sin^2 \theta \left[ 
(\partial_r \Omega)^2 + \frac{(\partial_{\theta} \Omega)^2}{r^2} \right] \, dV \ , \\
  &=& \int \eta r^2 \sin^2 \theta ( \ve{\nabla} \Omega)^2 \, dV \ . 
\eeqn

We note that, consistent with the nonrelativistic MHD approximation, the 
electric field energy $E^2/8\pi$ is not included in Eq.~(\ref{eq:Engterms}) 
and the angular momentum of the electromagnetic field is not included in 
Eq.~(\ref{eq:angJ})~\cite{landau84}.  

The motivation for monitoring the conservation equations during the evolution 
is twofold: physically, evaluating the individual terms enables us to track 
how the initial rotational energy and angular momentum in the fluid are 
transformed and/or dissipated; computationally, monitoring how well the 
conservation equations are satisfied provides a check on the numerical 
integration scheme.

\subsection{Build up of Small-Scale Structure: Phase Mixing}
\label{sec:sssNet}

In the absence of viscosity, the MHD equations~(\ref{eq:B1}) and~(\ref{eq:MHD1}) 
become
\beqn
  \partial_t B^{\phi} &=& B^j(0) \partial_j \Omega \ , \label{eq:B2}  \\
  \partial_t \Omega &=& \frac{1}{4\pi \rho r^2\sin^2 \theta}
B^j(0) \partial_j (r^2 \sin^2 \theta B^{\phi}) \ .
\label{eq:MHD2}
\eeqn
These equations have a peculiar property that will induce the growth of 
small-scale fluctuations in the fluid's angular motion and toroidal field.
To study this, we first notice that the 2+1 MHD equations can be reduced to a 
1+1 problem if we can 
find a coordinate transformation $u=u(r,\theta)$, $v=v(r,\theta)$ such that 
\beq
  B^j(0) \frac{\partial}{\partial x^j} = \kappa(u,v) 
\frac{\partial}{\partial u} \ .
\label{eq:coords}
\eeq
If such coordinates exist, the MHD equations~(\ref{eq:B2}) and~(\ref{eq:MHD2}) 
reduce to the following 1+1 form:
\beqn
  \partial_t B^{\phi} &=& \kappa(u,v) \partial_u \Omega \ ,
\label{eq:B3} \\
  \partial_t \Omega &=& \frac{1}{4\pi \rho r^2\sin^2 \theta} 
\kappa(u,v) \partial_u (r^2 \sin^2 \theta B^{\phi}) \ .
\label{eq:MHD3}
\eeqn
Here $r$ and $\theta$ are treated as implicit functions of $u$ and $v$. Applying 
Eq.~(\ref{eq:coords}) to $u$ and $v$, we obtain the conditions that must be 
satisfied by $u$ and $v$:
\beqn
  B^j(0) \frac{\partial u}{\partial x^j} &=& \kappa 
\label{eq:kappa} \ , \\ 
  B^j(0) \frac{\partial v}{\partial x^j} &=& 0 \ . \label{eq:veq}
\eeqn

Such coordinates indeed exist and are easy to find. Equation~(\ref{eq:veq}) requires 
that $v$ is constant along a given poloidal field line. Since the field lines 
do not cross and cover the volume insider the star, $v$ is a valid coordinate,  
labeling each field line. One way to find 
$v$ is by way of the vector potential. The covariant $\phi$-component of the vector 
potential is defined as
\beq
  A_{\phi} = r \sin \theta A^{\hat{\phi}} \ .
\eeq
It follows from $\ve{B}(0)=\ve{\nabla} \times (A^{\hat{\phi}} \ve{e}_{\hat{\phi}})$ 
that 
\beqn
  B^r(0) &=& \frac{\partial_{\theta} A_{\phi}}{r^2 \sin^2 \theta} \ , \\
  B^{\theta}(0) &=& -\frac{\partial_r A_{\phi}}{r^2 \sin^2 \theta} \ .
\eeqn
It follows that $B^j(0) \partial_j A_{\phi}=0$. Hence we can choose 
$v=f(A_{\phi})$, where $f$ is any function with a continuous first 
derivative. The coordinate $u$ is arbitrary as long as it 
is independent of $v$. The function $\kappa$ is then determined 
by Eq.~(\ref{eq:kappa}).

It is obvious that for the $l=1$ field ($\ve{B}(0)=B_0 \ve{e}_z$), 
the preferred coordinates are cylindrical coordinates 
($u=z=r\cos \theta$ and $v=\varpi=r \sin \theta$), for which 
$\kappa=B_0$. The MHD equations become 
\beqn
  \partial_t B^{\phi} &=& B_0 \partial_z \Omega \label{eq:BL1} \ , \\ 
  \partial_t \Omega &=& \frac{B_0}{4\pi \rho} \partial_z B^{\phi} 
\label{eq:MHDL1} \ ,
\eeqn
which can be solved analytically (see the next subsection).

The situation for the $l=2$ field is more complicated. The covariant $\phi$-component 
of the vector potential 
that generates the field is $A_{\phi}=B_0 (r^3/R) \sin^2 \theta \cos \theta$ 
[see Eq.~(\ref{eq:Al2})]. We choose 
\beqn
  u &=& -r^2 (3\cos^2 \theta -1) \ , \\
  v &=& r^3\sin^2 \theta \cos \theta \ .
\eeqn
The variable $u$ is chosen so that $\ve{\nabla} u \propto \ve{B}(0)$, 
hence $u \propto \Phi_m$, 
and $\ve{\nabla} u \cdot \ve{\nabla} v =0$. Equation~(\ref{eq:kappa}) 
gives
\beq
  \kappa = -\frac{2B_0 r^2}{R} (3\cos^2 \theta +1) \ .
\eeq
The MHD equations become 
\beqn
  \partial_t B^{\phi} &=& -\frac{2B_0 r^2}{R} (3\cos^2 \theta +1) \partial_u \Omega \\
  \partial_t \Omega &=& -\frac{2B_0}{4\pi \rho R} \left[ r^2 (3\cos^2 \theta+1) 
\partial_u B^{\phi}+B^{\phi} \right] \ .
\eeqn
Here $r$ and $\theta$ are regarded as implicit functions of $u$ and $v$.

The set of equations~(\ref{eq:B3}) and~(\ref{eq:MHD3}) is effectively a 1+1 
system, with $v$ serving as an effective ``mode'' number. There is no 
coupling between 
different values of $v$. The toroidal field $B^{\phi}$ and angular velocity 
$\Omega$ oscillate independently along each poloidal field 
line. The oscillation frequencies 
are determined by the boundary condition $B^{\phi}(r=R)=0$ and are in 
general different on different field lines. As a result, the toroidal 
field and angular velocity 
on different poloidal field lines will lose coherence. Small-scale 
structure will gradually build up. This is commonly referred to 
{\em phase mixing}~\cite{heyvaerts83,spruit99}.
In the next subsection, we demonstrate this phenomenon 
quantitatively via an analytic solution.

\subsection{Analytic Solution for the $\ve{l=1}$ field}
\label{sec:analsol}

Combining Eqs.~(\ref{eq:BL1}) and~(\ref{eq:MHDL1}), we obtain 
\beq
  \partial_t^2 B^{\phi} = v_A^2 \partial_z^2 B^{\phi} \ ,
\label{eq:wave}
\eeq
where $v_A=B_0/\sqrt{4\pi \rho}$ is the Alfv\'en velocity associated with 
the poloidal field. This is a simple wave equation; a similar equation 
arises for $\Omega$. 
We are interested in the solution in which $\Omega$ is symmetric under 
$z \rightarrow -z$ and $B^{\phi}=0$ at $t=0$. The general solution 
that satisfies the boundary condition $B^{\phi}(r=R)=0$ is 
\beqn
  B^{\hat{\phi}}(t;\varpi,z) &=& B_0 \left( \frac{\varpi}{R} \right)
 \sum_{n=1}^{\infty} 
C_n(\varpi) \sin [ k_n(\varpi)z] \times \cr & & \cr
& & \ \ \ \  \sin [\omega_n(\varpi) t] \ , 
\label{eq:analBphi} 
\eeqn
\newpage
\beqn
  \Omega(t;\varpi,z) &=& \Omega(0;\varpi,z) + \frac{v_A}{R} 
\sum_{n=1}^{\infty} C_n(\varpi) \cos [ k_n(\varpi)z] \times \cr & & \cr
 & & \{ 1-\cos [ \omega_n(\varpi) t] \} \ , \label{eq:analOmega}
\eeqn
where 
\beq
  k_n(\varpi) = \frac{n \pi}{\sqrt{R^2-\varpi^2}} \  \ , \ \  
  \omega_n(\varpi) = v_A k_n(\varpi) \ . 
\label{eq:omegan}
\eeq 
The functions $C_n(\varpi)$ are determined by the initial differential
angular velocity distribution $\Omega(0;\varpi,z)$: 
\beq
  C_n(\varpi) = \frac{2R}{\omega_n \sqrt{R^2-\varpi^2}}
\int_0^{\sqrt{R^2-\varpi^2}} [\partial_z \Omega(0;\varpi,z)] \sin k_n z\, dz \ .
\label{eq:Cn}
\eeq

We see clearly from Eq.~(\ref{eq:omegan}) that the characteristic frequency 
on each field line is different. Incoherence and small-scale structure 
will build up after a
certain time. To be precise, for a given $\varpi$ and a small length scale $\Delta L$, 
the eigenfrequency $\omega_n(\varpi)$ changes on this length scale by an amount 
\beq
  \Delta \omega_n = \frac{n \pi v_A \varpi}{(R^2-\varpi^2)^{3/2}} \Delta L \ .
\eeq
The field $B^{\hat{\phi}}$ and angular velocity $\Omega$ will lose coherence 
over $\Delta L$ on a timescale 
\beq
  t_{\rm coh} \approx \frac{2\pi}{\Delta \omega_n} = 
\frac{2 R^2}{n \varpi \Delta L} \left( 1 - \frac{\varpi^2}{R^2}\right)^{3/2} t_A \ ,
\eeq
where the Alfv\'en time $t_A=R/v_A$. 
For $\varpi=R/2$, $\Delta L = R/10$ and $n=1$, we have $t_{\rm coh} \approx 
26 t_A$. 
The phase mixing will eventually lead to 
chaotic-like motion over any finite radial scale.

The solution for the $l=2$ case must be obtained numerically. We postpone a 
discussion of this case, and the effects of viscous damping, for the fully 
relativistic treatment.

\section{General Relativistic MHD}
\label{sec:Rel}

\subsection{3+1 Decomposition}

In this subsection, we briefly summarize the general relativistic MHD formulation 
presented in Ref.~\cite{baumgarte03}. Hereafter, we adopt geometrized units 
and set $c=G=1$.

We start with the standard 3+1 decomposition of the metric: 
\beq ds^2=-\alpha^2 dt^2+\gamma_{ij} (dx^i+\beta^i dt)
(dx^j+\beta^j dt) \ . \eeq 
The spatial metric $\gamma_{ab}$ is related to the full metric $g_{ab}$ by 
\beq
  \gamma_{ab}=g_{ab}+ n_a n_b \ ,
\eeq
where $n^a$ is the unit normal vector $n_a = - \alpha \nabla_a t$ to the spatial 
slices. Next we decompose the Faraday tensor as 
\beq
  F^{ab}=n^a E^b - n^b E^a + \epsilon^{abc} B_c  
\label{eq:Fab}
\eeq
so that $E^a$ and $B^a$ are the electric and magnetic fields observed by a normal 
observer $n^a$. Both fields are purely spatial, i.e.\ 
\beq
  E^a n_a = 0 = B^a n_a \ .
\eeq
The three-dimensional Levi-Civita tensor is defined by 
\beqn 
\epsilon^{abc} &=& \epsilon^{abcd} n_d  \ , \\
\epsilon^{abcd} &=& -\frac{1}{\sqrt{-g}} [a\ b\ c\ d] \ ,
\eeqn
where $[a\ b\ c\ d]$ is the antisymmetrization symbol.
We also decompose the current four-vector $\mcal{J}^a$ as
\beq \mcal{J}^a = n^a \rho_e + J^a \ , \eeq
where $\rho_e$ and $J^a$ are the charge and current density as observed by 
a normal observer $n^a$. Note that $J^a$ is purely spatial.

With these definitions, Maxwell's equations $\nabla_b F^{ab}=4\pi \mcal{J}^a$ and 
$\nabla_{[a}F_{bc]}=0$ take the following 3+1 form:
\beqn
  D_i E^i &=& 4\pi \rho_e \ , \\
  \partial_t E^i &=& \epsilon^{ijk} D_j (\alpha B_k) 
  -4\pi \alpha J^i + \alpha K E^i + \pounds_{\beta} E^i \ , \ \ \ \  \\ 
  D_i B^i &=& 0 \label{eq:relDB0} \ ,  \\
  \partial_t B^i &=& -\epsilon^{ijk} D_j (\alpha E_k) 
  + \alpha K B^i + \pounds_{\beta} B^i  \ ,
\label{eq:faraday}
\eeqn
where $K$ is the trace of the extrinsic curvature, $D_i$ is the 
covariant derivative compatible with $\gamma_{ij}$, and $\pounds$ denotes 
a Lie derivative.

It follows from $\nabla_{[a}F_{bc]}=0$ that $F_{ab}$ can be expressed in 
terms of a vector potential 
\beq
  F_{ab}=\partial_a \mcal{A}_b-\partial_b \mcal{A}_a \ .
\eeq
We decompose $\mcal{A}_a$ according to 
\beq \mcal{A}_a=\Phi_e n_a + A_a \ , \eeq
where the three-vector potential $A^a$ as observed by a normal observer 
$n^a$ is purely spatial ($A^a n_a=0$).
The relationship between $\Phi_e$, $A_k$, $E^i$ and $B^j$ are given by 
\beqn
  \partial_t A_i &=& -\alpha E_i - \partial_i (\alpha \Phi_e) 
 + \pounds_{\beta} A_i \ , \\ 
  B^i &=& \epsilon^{ijk} A_{k,j} \label{eq:BviaA} \ .
\eeqn

In the ideal MHD limit, the fluid is assumed to have perfect conductivity, which 
is equivalent to the condition that the electric field vanish in the fluid's 
rest frame. The relation between the electric and magnetic field is given by 
\beq 
\alpha E_i = -\epsilon_{ijk} (v^j + \beta^j) B^k \ , \label{eq:E-MHD}
\eeq
where $v^j=u^j/u^t$. In the Newtonian limit, the above equation reduces to 
the familiar form $\ve{E}=-\ve{v}\times \ve{B}$.

In the ideal MHD limit, the Faraday's law~(\ref{eq:faraday}) 
becomes 
\beq 
\partial_t \mcal{B}^i = \partial_j (v^i \mcal{B}^j-v^j \mcal{B}^i) \ ,
\label{eq:relB0}
\eeq
where $\mcal{B}^i=\sqrt{\gamma} B^i$. Here $\gamma$ is the determinant of the 
spatial metric $\gamma_{ij}$. Equation~(\ref{eq:relB0}) is the relativistic 
version of Eq.~(\ref{eq:B0}).

To derive the relativistic Navier-Stokes equation in the ideal MHD limit, we 
separate the stress-energy tensor into a fluid part and an electromagnetic part:
\beq T^{ab}=T^{ab}_{\rm{fluid}}+T^{ab}_{\rm{em}} \ , \eeq
where 
\beqn
  T^{ab}_{\rm{fluid}} &=& T^{ab}_{\rm{p}}+T^{ab}_{\rm{vis}} \\
  T^{ab}_{\rm{p}} &=& \rho h u^a u^b + P g^{ab} \\ 
  T^{ab}_{\rm{em}} &=& \frac{1}{4\pi} \left( F^{ac} {F^b}_c 
   	-\frac{1}{4} g^{ab} F_{cd} F^{cd} \right) \ .
\eeqn
Here $h=1+\epsilon+P/\rho$ is the specific enthalpy, $\rho \epsilon$ is the 
internal energy density, and $T_p^{ab}$ is the stress-energy tensor for a perfect 
fluid. An incompressible fluid may be regarded as a 
gamma-law equation of state $P=(\Gamma-1) \rho \epsilon$ in the limit 
$\Gamma \rightarrow \infty$. Hence we have $\epsilon = 0$ for an incompressible 
fluid.
The viscous stress-energy tensor $T^{ab}_{\rm{vis}}$ is given by [see, 
e.g.~\cite{MTW}, Eq.~(22.16a)]
\beq
  T^{ab}_{\rm{vis}} = -2\eta \sigma^{ab}-\zeta P^{ab} \nabla_c u^c  \ .
\label{eq:visTab}
\eeq
The projection tensor $P^{ab}$ and the shear tensor $\sigma^{ab}$ are 
defined as
\beqn 
  P^{ab} &=& g^{ab} + u^a u^b \\
  \sigma^{ab} &=& \frac{1}{2} \left( \nabla_c u^a P^{cb} + \nabla_c u^b P^{ca} 
\right) -\frac{1}{3} P^{ab} \nabla_c u^c \ .
\label{eq:sigmaab}
\eeqn
For an incompressible fluid, $\nabla_c u^c=0$.
The relativistic Navier-Stokes equation is computed from the spatial 
components of the equation $\nabla_a T^{ab}=0$. The result is 
\beqn
  \partial_t (\sqrt{\gamma}\, S_i)+\partial_j(\sqrt{\gamma}\, S_i v^j) 
= -\alpha \sqrt{\gamma}\, \left( \partial_i P + \frac{S_a S_b}{2\alpha S^t} 
\partial_i g^{ab}\right) & &   \cr & & \cr
+ \alpha \sqrt{\gamma}\, F_{ia} \mcal{J}^a 
+ 2 \partial_b (\alpha \sqrt{\gamma}\, \eta \sigma_i^b)+\eta \sigma_{bc} 
\partial_i g^{bc} \ , \ \ \ \ \ \ \ \  & &
\label{eq:relEuler0}
\eeqn
where 
\beq S_a = \rho h \alpha u^t u_a \ . \label{eq:Sa} \eeq
Lorentz's force law implies 
\beq 
4\pi \mcal{J}^a = \nabla_b F^{ab} = \frac{1}{\sqrt{-g}} 
\partial_b (\sqrt{-g} F^{ab}) \ .
\eeq
Hence we have 
\beq 
\alpha \sqrt{\gamma}\, F_{ia} \mcal{J}^a = \frac{1}{4\pi} 
F_{ia} \partial_b (\alpha \sqrt{\gamma}\, F^{ab}) \ .
\label{eq:Lorentz}
\eeq
Equations~(\ref{eq:relEuler0}) and~(\ref{eq:Lorentz}) comprise the 
relativistic version of Eq.~(\ref{eq:MHD0}).

\subsection{Basic Equations}

The metric for a uniform density, spherical star is given by (see, 
e.g.~\cite{MTW}, Box~23.2)
\beq
  ds^2 = -\alpha^2(r) dt^2 + \frac{dr^2}{\lambda^2(r)}+r^2 (d\theta^2 
+ \sin^2 \theta d\phi^2) \ ,
\label{bmetric}
\eeq
where
\beq 
 \alpha(r) = \left \{ \begin{array}{ll} 
\frac{3}{2}\sqrt{1-\frac{2M}{R}}-\frac{1}{2} 
\sqrt{1-\frac{2Mr^2}{R^3}} &  (r \leq R)  \\
 \sqrt{1-\frac{2M}{r}} & (r \geq R) \end{array} \right. \ ,
\eeq
and 
\beq
  \lambda(r) = \left \{ \begin{array}{ll} 
  \sqrt{1-\frac{2Mr^2}{R^3}} & (r \leq R) \\ 
  \sqrt{1-\frac{2M}{r}} & (r \geq R) \end{array} \right. \ .
\eeq
The interior pressure is 
\beq
  P(r) = \frac{\lambda(r)-\alpha_R}{3\alpha_R-\lambda(r)} \rho \ , 
\label{eq:bP}
\eeq
where 
\beq
  \alpha_R=\alpha(r=R)=\sqrt{1-\frac{2M}{R}} \ .
\eeq

In the presence of slow rotation, small magnetic field and viscosity, 
the diagonal components of the background metric has a correction 
of order $\Omega^2$, $|\ve{B}|^2$ and $\eta |\ve{\nabla} \Omega|$, which will be 
neglected since they are of higher nonlinear order. Rotation induces 
the off-diagonal component $g_{t\phi}$, which corresponds to the 
dragging of inertia frames. We write 
\beq
  g_{t\phi} = \beta_{\phi} = -r^2 \sin^2 \theta \, \omega(r,\theta) \ ,
\eeq
where $\omega(r,\theta)$ is the angular frequency of the zero 
angular momentum observer (ZAMO) observed by an inertial observer 
at infinity.
In order of magnitude, $\omega/\Omega \sim M/R$. 
We adopt the Cowling approximation, which assumes that the background 
metric is fixed, i.e.\ $\partial_t g_{\mu \nu}=0$. We can justify 
this approximation by the following argument. First, 
we can use the gauge degrees of freedom to freeze the lapse and 
shift: 
$\partial_t \alpha = \partial_t \beta_i=0$. Second, the evolution of the 
spatial metric $\gamma_{ij}$ is given by the ADM equations~\cite{york79}
\beqn
  \partial_t \gamma_{ij} &=& -2\alpha K_{ij}+D_i \beta_j + D_j \beta_i \ , \\
  \partial_t K_{ij} &=& -D_i D_j \alpha + \alpha (R_{ij} 
-2K_{ik} {K^k}_j + K K_{ij})  \cr  & & \cr
& & - 8\pi \alpha \left[ S_{ij}-\frac{1}{2}\gamma_{ij} 
(S-\rho_s) \right] + \beta^k D_k K_{ij} \cr & & \cr
  & & +K_{ik} D_j \beta^k + K_{kj} D_i \beta^k \ ,
\eeqn
where $K_{ij}$ is the extrinsic curvature, $R_{ij}$ is the three-dimensional 
Ricci tensor, $K={K^j}_j$, and 
\beqn
  \rho_s &=& n_{\mu} n_{\nu} T^{\mu \nu} \ , \\ 
  S_{ij} &=& \gamma_{ik} \gamma_{jm} T^{km} \ \ \ , \ \ \ S={S^j}_j \ ,
\eeqn
where $T^{\mu \nu}$ is the total stress-energy tensor. We assume that 
the star is in hydrostatic quasi-equilibrium at $t=0$. The time derivatives 
$\partial_t \gamma_{ij}$ and $\partial_t K_{ij}$ 
come from the changes in the 
matter and $B$-field sources term
$\delta \rho_s$, $\delta S_{ij}$ and $\delta S$, which are all of the order 
$\Omega^2$, $|\ve{B}|^2$ and $\eta |\ve{\nabla} \Omega|$. This means that the 
deviation of $\gamma_{ij}$ from its initial value will remain higher order.
Hence our gauge conditions does not introduce a first order correction 
to the spatial metric in later times, and the Cowling approximation is valid.

As in the Newtonian case, we assume that the system is 
axisymmetric and $v^r(t;r,\theta)=v^{\theta}(t;r,\theta)=0$, 
$v^{\phi}(t;r,\theta)=\Omega(t;r,\theta)$, $B^{\phi}(0;r,\theta)=0$. It follows from 
Eq.~(\ref{eq:relB0}) that $\partial_t B^r=\partial_t B^{\theta}=0$. Hence 
the poloidal field remains unchanged, just like the Newtonian case. Denote 
the initial magnetic field 
\beq
\ve{B}(0;r,\theta)=B^r(0;r,\theta) \ve{e}_r + B^{\theta}(0;r,\theta) \ve{e}_{\theta} \ ,
\eeq
where $\ve{e}_r=\partial/\partial r$ and 
$\ve{e}_{\theta}=\partial/\partial \theta$.
Taking the $\phi$-component of Eq.~(\ref{eq:relB0}), we have 
\beqn
  \partial_t B^{\phi} &=& \frac{1}{\sqrt{\gamma}} \partial_j 
(\Omega \sqrt{\gamma} B^j) \cr
 &=& \Omega \frac{1}{\sqrt{\gamma}}\partial_j (\sqrt{\gamma}B^j) + 
B^j \partial_j \Omega \cr  & & \cr
 &=& B^j(0) \partial_j \Omega  \ ,
\label{eq:relB01}
\eeqn
where we have used Eq.~(\ref{eq:relDB0}) to obtain the last equality.

The $\phi$-components of Eqs.~(\ref{eq:relEuler0}) and~(\ref{eq:Lorentz}) 
simplify to
\beq
  \partial_t S_{\phi} = \frac{1}{4\pi \sqrt{\gamma}} F_{\phi b} 
\partial_c (\alpha \sqrt{\gamma} F^{bc}) + 2 \partial_b (\alpha \sqrt{\gamma} 
\eta \sigma_{\phi}^b) \ .
\eeq
It follows from $v^{\phi}=\Omega$ and $u^{\alpha} u_{\alpha}=-1$ that
\beqn
  u^t &=& \frac{1}{\alpha} + O(\Omega^2) \ , \ \ u^{\phi} = \frac{\Omega}{\alpha}+ 
O(\Omega^3) \ , \\
  u_{\phi} &=& g_{\phi \mu} u^{\mu} = \frac{r^2 \sin^2 \theta}{\alpha} 
(\Omega-\omega) + O(\Omega^3) \ .
\eeqn
Using Eq.~(\ref{eq:Sa}), we obtain
\beqn
  S_{\phi} &=& \frac{\rho + P}{\alpha}\, r^2 \sin^2 \theta\, 
(\Omega-\omega) + O(\Omega^3)  \ . 
\eeqn
Straightforward calculations from Eqs.~(\ref{eq:Fab}) 
and~(\ref{eq:E-MHD}) using the metric~(\ref{bmetric}) yield, to the leading 
order,  
\beqn
  F_{\phi b} \partial_c (\alpha \sqrt{\gamma} F^{bc}) &=& 
  \frac{r^2 \sin \theta}{\lambda} B^j(0) \partial_j ( \alpha r^2 \sin^2 
\theta B^{\phi}) \cr 
 & & - \frac{r^4 \sin^3 \theta}{\alpha \lambda} \gamma_{ij} 
B^i(0) B^j(0) \partial_t \Omega \ , \ \ \ \ \ \ \ 
\eeqn 
where we have imposed our gauge condition $\partial_t \beta_i=0$ by 
dropping a term involving $\partial_t \omega$.
The viscosity term is [from Eq.~(\ref{eq:sigmaab})] 
\vskip 2.5cm
\beqn
  2 \partial_b (\alpha \sqrt{\gamma} \eta {\sigma^b}_{\phi}) &=& 
\partial_r (\eta \lambda r^4 \sin^3 \theta \partial_r \Omega) 
+ \partial_{\theta} \left( \eta \frac{r^2 \sin^3 \theta}{\lambda} 
\partial_{\theta} \Omega \right) \cr  & & \cr
& &  + O(\eta \Omega^2) \ .
\eeqn
Gathering all the terms, we obtain
\beqn
  \left[ 1 + \frac{\gamma_{ij} B^i(0) B^j(0)}{4\pi (\rho+P)} \right] 
\partial_t \Omega = \frac{\alpha B^j(0) \partial_j (\alpha r^2 
\sin^2 \theta B^{\phi})}{4\pi (\rho+P) r^2 \sin^2 \theta}
  & & \cr
 + \frac{\alpha \lambda}{(\rho+P) r^4 \sin^3 \theta} 
\left[\partial_r (\eta r^4 \lambda \sin^3 \theta \partial_r \Omega) 
\right. & &  \cr
 \left. + \partial_{\theta} \left( \eta \frac{r^2 \sin^3 \theta}{\lambda} 
\partial_{\theta} \Omega \right) \right] \ \ \ \ \ \ \ \ \ \ \ \ \ \ 
\ \ \ \ \ \   & & 
\eeqn
The term on the left-hand side, 
$\gamma_{ij} B^i(0) B^j(0)/4\pi (\rho+P) \sim v_A^2 \sim E_{\rm mag}/M$, 
is assumed to be small. Hence it can be neglected. We also note that 
to the leading order in our adopted gauge, the frame dragging frequency 
$\omega$ does not 
enter into the evolution equations. It follows 
from Eq.~(\ref{eq:bP}) that
\beq
  \rho + P(r) = \rho \frac{\alpha_R}{\alpha(r)} \ .
\eeq
Combining the results, we obtain the relativistic version of Eqs.~(\ref{eq:B1}) 
and~(\ref{eq:MHD1}): 
\beqn
  \partial_t B^{\phi} &=& B^j(0) \partial_j \Omega \label{eq:relB1} \\
  \partial_t \Omega &=& \frac{\alpha^2}{\alpha_R} \frac{1}{4\pi \rho r^2 
\sin^2 \theta} B^j(0) \partial_j (\alpha r^2 \sin^2 \theta B^{\phi})  \cr
 & & \cr & & + (\partial_t \Omega)_{\rm{vis}} \label{eq:relMHD1} \ ,
\eeqn
where 
\beq 
  (\partial_t \Omega)_{\rm{vis}} = \frac{\alpha^2 \lambda}{\alpha_R} 
\left[ \frac{1}{r^4} \partial_r (\nu \lambda r^4 \partial_r \Omega) 
+ \frac{1}{\lambda r^2 \sin^3 \theta} \partial_{\theta} (\nu \sin^3 
\theta \partial_{\theta} \Omega) \right] \ .
\label{eq:relvisterm}
\eeq
Here we again set $\nu=\eta/\rho$. The condition that the toroidal field 
cannot be carried
outside the star gives the boundary conditions $B^{\phi}=0$ at $r=R$, which 
also implies [from Eq.~(\ref{eq:relB1})] $B^j(0) \partial_j \Omega=0$ at $r=R$. 
In the Newtonian limit, with $M/R \ll 1$, Eqs.~(\ref{eq:relB1})--(\ref{eq:relvisterm}) 
reduce to the Newtonian MHD equations~(\ref{eq:B1})--(\ref{eq:vis}).

\subsection{Initial Magnetic Field}
\label{sec:RelIMF}

The two sets of poloidal field functions we discussed in Section~\ref{sec:IMF} cannot 
be used in the relativistic case because they do not satisfy the 
relativistic Maxwell equation $D_i B^i(0)=0$. 
We want to choose two sets of initial field which will reduce to the ones 
we discussed in Section~\ref{sec:IMF} in the Newtonian limit.
It is easy to prove that any axisymmetric poloidal field can be generated by 
a vector potential of the form $A_i = A_{\phi} {\delta_{i}}^{\phi}$. Hence we 
can specify $A_{\phi}$ and then compute $B^j(0)$ from Eq.~(\ref{eq:BviaA}). 
This guarantees that the constraint equation $D_i B^i(0)=0$ is satisfied.

The simplest way to generalize the Newtonian initial fields 
is to use the same $A_{\phi}$ as in the Newtonian cases [Eqs.~(\ref{eq:Al1}) 
and~(\ref{eq:Al2})]. 
Hence the ``$l=1$'' field is generated by
\beq A_{\phi} = \frac{1}{2} B_0 r^2 \sin^2 \theta \ . \eeq
The corresponding poloidal field is given by Eq.~(\ref{eq:BviaA}), yielding
\beq 
  B^r = \lambda B_0 \cos \theta  \ , \ \ B^{\theta} = -\lambda \frac{B_0}
{r} \sin \theta \ . \label{eq:IMFrel}
\eeq
We define cylindrical coordinates $\varpi = r \sin \theta$, $z=r \cos \theta$. 
The cylindrical components of the poloidal field are
\beq
  B^{\varpi}=0 \ , \ \ B^z=\lambda B_0 \ .
\eeq
We see that the poloidal field is still in the $z$ direction. However, its 
amplitude 
\beqn
  |\ve{B}(0)| &=& \sqrt{\gamma_{ij} B^i(0) B^j(0)} \cr
  &=& B_0 \sqrt{1-\frac{2M\varpi^2}{R^3}}
\eeqn
decreases with increasing $\varpi$.

The generalized ``$l=2$'' field is generated by 
\beq A_{\phi} = B_0 \frac{r^3}{R} \sin^2 \theta \cos \theta \ . \eeq
The poloidal field has components 
\beq
  B^r = \lambda B_0 \frac{r}{R} (3 \cos^2 \theta -1) \ , \ \ 
  B^{\theta} = -\frac{3 \lambda B_0}{R} \sin \theta \cos \theta \ .
\label{eq:IMFrel2}
\eeq
It follows that the shape of the field line of this $l=2$ field is the same as 
the Newtonian $l=2$ field shown in Fig.~\ref{fig:L2field}, even though the 
magnitude is different. 

\subsection{Conserved Integrals}

As in the Newtonian case, the relativistic MHD equations~(\ref{eq:relB1}) 
and~(\ref{eq:relMHD1}) admit two non-trivial integrals of motion, expressing 
the conservation of energy and angular momentum. The derivation is very 
similar to the Newtonian case, so we just state the result here.

The conserved energy contains a sum of three terms:
\beq
  E=E_{\rm{rot}}+E_{\rm{mag}}+E_{\rm{vis}} \ ,
\eeq
where 
\beqn
  E_{\rm{rot}} &=& \frac{1}{2} \int
\left( \frac{\rho+P}{\alpha}\Omega^2 r^2 \sin^2 \theta \right)
\frac{r^2 \sin \theta}{\lambda} dr d\theta d\phi , \ \ \ \ \ \  \\
  E_{\rm{mag}} &=& \int \alpha \frac{(B^{\hat{\phi}})^2}{8\pi}
\frac{r^2 \sin \theta}{\lambda} dr d\theta d\phi \ , \\
  E_{\rm{vis}} &=& \int_0^t \dot{E}_{\rm{vis}}(t') dt' \ ,
\eeqn
and where 
\beqn
  \dot{E}_{\rm{vis}} &=& \int \eta r^2 \sin^2 \theta \left[ \lambda^2 
(\partial_r \Omega)^2 + \frac{(\partial_{\theta} \Omega)^2}{r^2} \right] 
\frac{r^2 \sin \theta}{\lambda} dr d\theta d\phi \ \ \ \ \ \ \ \ \\ 
  &=& \int \eta r^2 \sin^2 \theta g^{ij} \nabla_i \Omega \nabla_j \Omega 
 \frac{r^2 \sin \theta}{\lambda} dr d\theta d\phi \ .
\eeqn

The conserved angular momentum is 
\beq
  J = \int \frac{\rho+P}{\alpha} (\Omega r^2 \sin^2 \theta) 
\frac{r^2 \sin \theta}{\lambda} dr d\theta d\phi \ .
\eeq

\subsection{Special Coordinates and Phase Mixing}
\label{sec:sssRel}

As in the Newtonian case, in the absence of viscosity,
Eqs.~(\ref{eq:relB1}) and~(\ref{eq:relMHD1}) 
can be brought into 1+1 form by the coordinate transformation 
$u=u(r,\theta)$, $v=v(r,\theta)$ that satisfies
\beqn
  B^j(0) \frac{\partial u}{\partial x^j} &=& \kappa(u,v) \ , \label{eq:relkappa} \\
  B^j(0) \frac{\partial v}{\partial x^j} &=& 0 \ .
\eeqn
In these new coordinates, Eqs~(\ref{eq:relB1}) and~(\ref{eq:relMHD1}) 
become (without viscosity)
\beqn
  \partial_t B^{\phi} &=& \kappa \partial_u \Omega \label{eq:relB4} \\ 
  \partial_t \Omega &=& \frac{\alpha^2}{\alpha_R} \frac{\kappa}{4\pi 
\rho r^2 \sin^2 \theta} \partial_u (\alpha r^2 \sin^2 \theta B^{\phi}) \ .
\label{eq:relMHD4}
\eeqn
It follows from Eq.~(\ref{eq:BviaA}) that when $A_i=A_{\phi} 
{\delta_{i}}^{\phi}$, 
\beq
  B^r = \frac{1}{\sqrt{\gamma}} \partial_{\theta} A_{\phi} \ , \ \ \ \ \ 
  B^{\theta} = -\frac{1}{\sqrt{\gamma}} \partial_r A_{\phi} \ .
\eeq
Hence we have $B^i \partial_i A_{\phi}=0$. We can choose $v=f(A_{\phi})$ 
as in the Newtonian case, where $f$ is an arbitrary function with 
continuous first derivative. Since we use the same $A_{\phi}$ as the 
Newtonian case in constructing the initial magnetic field, we can use the 
same $u$ and $v$ to simplify the MHD equations.

In particular, we use cylindrical coordinates defined by $\varpi = r \sin \theta$ 
and $z=r \cos \theta$ for the $l=1$ field. In these coordinates, the 
MHD equations become 
\beqn
  \partial_t B^{\phi} &=& \lambda B_0 \partial_z \Omega \label{eq:relB5} \\
  \partial_t \Omega &=& \frac{\alpha^2 \lambda}{\alpha_R} \left( 
\frac{B_0}{4\pi \rho} \right) \partial_z (\alpha B^{\phi}) \ .
\label{eq:relMHD5}
\eeqn

For the $l=2$ field, we choose
\beqn
  u &=& -r^2 (3\cos^2 \theta -1) \ , \\
  v &=& r^3\sin^2 \theta \cos \theta \ .
\eeqn
It follows from Eq.~(\ref{eq:relkappa}) that 
\beq
  \kappa=B^j(0)\partial_j u = -\frac{2\lambda B_0 r^2}{R} (3\cos^2 \theta +1) \ .
\eeq
The MHD equations become
\beqn
  \partial_t B^{\phi} &=& -\frac{2\lambda B_0 r^2}{R} (3\cos^2 \theta +1) 
\partial_u \Omega \\
  \partial_t \Omega &=& - \left( \frac{\alpha^2 \lambda}{\alpha_R}\right)
\frac{2 B_0}{4\pi \rho R} [ r^2 (3\cos^2 \theta+1)
\partial_u (\alpha B^{\phi}) \cr & & \cr 
& & +\alpha B^{\phi} ] \ .
\eeqn
Here $r$ and $\theta$ are regarded as implicit functions of $u$ and $v$.

As discussed in Section~\ref{sec:sssNet}, Eqs.~(\ref{eq:relB4}) and~(\ref{eq:relMHD4}) 
imply that there is no coupling of $B^{\phi}$ and $\Omega$ between different 
poloidal field lines. The toroidal field and angular velocity will oscillate 
in general at different characteristic frequencies along different field lines. 
The angular velocity flow and toroidal field pattern will become irregular 
as a result. We will demonstrate this phase mixing effect 
in Section~\ref{sec:numerics} when we present our numerical results.

\section{Nondimensional Formulation}
\label{sec:nondim}

Before numerically integrating the MHD equations, it is convenient to introduce 
the following nondimensional variables:
\beqn
  \hat{r} &=& \frac{r}{R} \ , \\
  \hat{t} &=& \frac{B_0 t}{R\sqrt{4\pi \rho}} = \frac{t}{t_A} \ , \\
  \hat{B} &=& \frac{\alpha B^{\phi}}{\Omega_c \sqrt{4\pi \rho}}
  = \frac{\alpha B^{\hat{\phi}}}{\Omega_c r \sin \theta \sqrt{4\pi \rho}} \ , \\
  \hat{B}^j(0) &=& \frac{B^j(0)}{B_0} \ \ \ \ \ , \ \ \ j=r,\ \theta \ ,\\
  \hat{\Omega} &=& \frac{\Omega}{\Omega_c} \ , \\
  \hat{\nu} &=& \frac{\nu}{R} \left( \frac{\sqrt{4\pi \rho}}{B_0} \right) \ , \\
  \hat{E} &=& \frac{E}{M (R\Omega_c)^2} \ , \\
  \hat{J} &=& \frac{J}{MR^2\Omega_c} \ ,
\eeqn
where the Alfv\'en time $t_A$ is defined as 
\beq
  t_A = \frac{R}{B_0/\sqrt{4\pi \rho}} \ .
\eeq
Here $\Omega_c$ is an arbitrary constant with a magnitude characteristic 
of the initial angular velocity.
In terms of these new variables, Eqs.~(\ref{eq:relB1}) and~(\ref{eq:relMHD1}) 
become
\beqn
  \partial_{\hat{t}} \hat{B} &=& \alpha \hat{B}^j(0) \partial_{\hat{j}} 
\hat{\Omega} \label{eq:relB2} \\
  \partial_{\hat{t}} \hat{\Omega} &=& \frac{\alpha^2}{\alpha_R} 
\frac{\hat{B}^j(0)}{\hat{r}^2 \sin^2 \theta} \partial_{\hat{j}} 
(\hat{r}^2 \sin^2 \theta \hat{B}) + (\partial_{\hat{t}} 
\hat{\Omega})_{\rm vis} \label{eq:relMHD2} \ , \ \ \ \ \ 
\eeqn
where $\partial_{\hat{j}}=R\partial_j$ and 
\beqn
  (\partial_{\hat{t}}\hat{\Omega})_{\rm vis} &=& \frac{\alpha^2 \lambda}
{\alpha_R \hat{r}^4 \sin^3 \theta} \left[ \partial_{\hat{r}}(\hat{\nu} 
\hat{r}^4 \sin^3 \theta \partial_{\hat{r}} \hat{\Omega}) \right. \cr
& & \left. + \partial_{\theta} 
\left( \hat{\nu} \frac{\hat{r}^2 \sin^3 \theta}{\lambda} \partial_{\theta} 
\hat{\Omega} \right) \right]  \ .
\eeqn
Since the background pressure distribution is spherical, it is reasonable 
to assume that $\nu=\nu(r)$, hence 
\beqn
  (\partial_{\hat{t}}\hat{\Omega})_{\rm vis} &=& 
\frac{\alpha \lambda}{\alpha_R} \left \{ \hat{\nu} \lambda 
\partial_{\hat{r}}^2 \hat{\Omega} + \left[ \partial_{\hat{r}} (\nu \lambda) 
+ \frac{4}{\hat{r}}\nu \lambda \right] \partial_{\hat{r}} \hat{\Omega} 
 \right. \cr
  & & \left. + \frac{\hat{\nu}}{\lambda \hat{r}^2} (\partial_{\theta}^2 
\hat{\Omega} + 3 \cot \theta \partial_{\theta} \hat{\Omega}) \right \} \ .
\label{eq:visterm}
\eeqn
The Navier-Stokes equation and the regularity condition on the surface of the 
star require that the dynamic viscosity $\eta$ in general has to 
decrease to zero continuously as the surface of the star is approached. 
We adopt the simplest viscosity model in our numerical 
computations: we assume that $\nu=\eta/\rho$ is essentially constant in the 
interior, but rapidly drops 
to zero when approaching the surface of the star at $r=R$.

While the variable $\hat{B}$ is convenient for numerical evolution of the 
MHD equations, we also found it convenient to introduce another non-dimensional 
variable 
\beq
  \bar{B} = \frac{\hat{r} \sin \theta}{\alpha} \hat{B} = \left( \frac{B^{\hat{\phi}}}
{B_0} \right) \left( \frac{v_A}{\Omega_c R} \right) \ ,
\label{eq:Bbar}
\eeq
which is a measure of the strength of the toroidal magnetic field. Note that 
in this non-dimensional formulation, we do not need to specify the 
ratio $v_A/\Omega_c R$, which is assumed to be much greater than one so that 
$E_{\rm mag} \gg E_{\rm rot}$. We 
also do not need to specify the characteristic amplitude 
of the seed poloidal magnetic field $B_0$, which has to be small so that 
$E_{\rm mag} \ll W$ or equivalently, $v_A \ll 1$. These quantities are absorbed in our 
nondimensional variables.

In the next two subsections, we will further simplify the general equations 
above by inserting the specific forms of the two sets of initial 
magnetic field.

\subsection{Equations for the $\ve{l=1}$ Field}
\label{sec:ndl=1}

As discussed in Sections~\ref{sec:sssNet} and~\ref{sec:sssRel}, the 
cylindrical coordinates $\hat{\varpi}=\hat{r} \sin \theta$ and
$\hat{z}=\hat{r} \cos \theta$ are the preferred coordinates 
in the absence of viscosity. To impose the boundary condition 
at $\hat{r}=1$, it is convenient to introduce a new variable 
$\hat{u}=\hat{z}/\sqrt{1-\hat{\varpi}^2}$ in place of $\hat{z}$. The boundary of 
the star is located at $\hat{u}=1$. We are interested in a solution
in which $\Omega$ is symmetric under the reflection $z \rightarrow -z$.
Hence the computation domain is $\hat{u} \in [0,1]$ and
$\hat{\varpi} \in [0,1]$. The MHD equations, in terms of 
$\hat{u}$ and $\hat{\varpi}$, take the form  
\beqn
  \partial_{\hat{t}} \hat{B} &=& \frac{\alpha \lambda}
{\sqrt{1-\hat{\varpi}^2}} \partial_{\hat{u}} \hat{\Omega} 
\label{eq:dtBl1} \\
  \partial_{\hat{t}} \hat{\Omega} &=& \frac{\alpha^2 \lambda}{\alpha_R
\sqrt{1-\hat{\varpi}^2}} 
\partial_{\hat{u}} \hat{B} + (\partial_{\hat{t}} \hat{\Omega})_{\rm vis} \ ,
\label{eq:dtWl1}
\eeqn
where $(\partial_{\hat{t}} \hat{\Omega})_{\rm vis}$ is obtained from 
Eq.~(\ref{eq:visterm}) and Eqs.~(\ref{eq:dromega})--(\ref{eq:dth2omega}).

The nondimensional energy and angular momentum are given by 
\beqn
  \hat{J} &=& 3 \alpha_R \int_0^1 d \hat{\varpi} \hat{\varpi}^3 
\sqrt{1-\hat{\varpi}^2} \int_0^1 d\hat{u} \frac{\hat{\Omega}}{\alpha^2 \lambda}
\label{eq:nondJ} \\
  \hat{E} &=& \hat{E}_{\rm rot}(\hat{t}) + \hat{E}_{\rm mag}(\hat{t}) 
+ \int_0^{\hat{t}} dt' \dot{\hat{E}}_{\rm vis}(t') \ ,  
\eeqn
where 
\beqn
  \hat{E}_{\rm rot} &=& \frac{3}{2} \int_0^1 d \hat{\varpi} \hat{\varpi}^3 
\sqrt{1-\hat{\varpi}^2} \, \hat{\epsilon}_1^{\rm rot}(\hat{\varpi}) \ , \\
  \hat{E}_{\rm mag} &=& \frac{3}{2} \int_0^1 d \hat{\varpi} \hat{\varpi}^3 
\sqrt{1-\hat{\varpi}^2} \, \hat{\epsilon}_1^{\rm mag}(\hat{\varpi}) \\
  \dot{\hat{E}}_{\rm vis} &=& -3 \int_0^1 d\hat{\varpi} \int_0^1 d\hat{z} 
\frac{\hat{\varpi}^3 \hat{\nu}}{\lambda} \left[ \lambda^2 
(\partial_{\hat{r}} \hat{\Omega})^2 + \frac{(\partial_{\theta} 
\hat{\Omega})^2} {\hat{r}^2} \right] \ \ \ \ \ \ \
\eeqn
The functions $\hat{\epsilon}_1^{\rm rot}(\hat{\varpi})$ and 
$\hat{\epsilon}_1^{\rm mag}(\hat{\varpi})$ are given by 
\beqn
  \hat{\epsilon}_1^{\rm rot}(\hat{\varpi}) &=&  \int_0^1 d\hat{u} \frac{\alpha_R}
{\alpha^2 \lambda} \hat{\Omega}^2 \ , \label{eq:e1rot} \\
  \hat{\epsilon}_1^{\rm mag}(\hat{\varpi}) &=& \int_0^1 d\hat{u} 
\frac{\hat{B}^2}{\alpha \lambda} \ . \label{eq:e1mag}
\eeqn
In the absence of viscosity, there is no coupling between different 
values of $\hat{\varpi}$. Hence the reduced energy and angular momentum functions 
\beqn
  \hat{\epsilon}_1(\hat{\varpi}) &=& \hat{\epsilon}_1^{\rm rot}(\hat{\varpi}) + 
\hat{\epsilon}_1^{\rm mag}(\hat{\varpi}) \label{eq:el1} \ , \\
  \hat{j}_1(\hat{\varpi}) &=& \int_0^1 d\hat{u} \frac{\hat{\Omega}}
{\alpha^2 \lambda} 
\label{eq:jl1}
\eeqn
are also conserved for each value of $\hat{\varpi}$. This can also be 
proved directly from Eqs.~(\ref{eq:dtBl1}) and~(\ref{eq:dtWl1}).

\subsection{Equations for the $\ve{l=2}$ Field}
\label{sec:ndl=2}

\begin{figure}
\includegraphics[width=6cm]{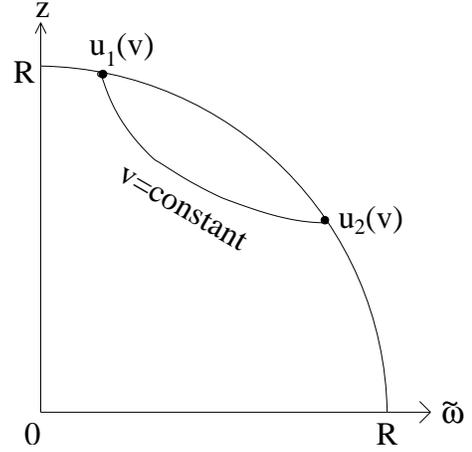}
\vskip 0.5cm
\caption{The points $u_1(v)$ and $u_2(v)$.} 
\label{fig:u1u2}
\end{figure}

In the absence of viscosity, the preferred coordinates 
are $u=-r^2 (3\cos^2 \theta - 1)$ and $v = r^3 \sin^2 \theta \cos \theta$. 
In numerical calculations, it is more convenient to introduce the following 
nondimensional variables
\beqn
  \hat{u} &=& \frac{u-u_1(v)}{u_2(v)-u_1(v)} \label{eq:uhat} \\
  \hat{v} &=& \frac{3 \sqrt{3}}{2} \hat{r}^3 \sin^2 \theta \cos \theta \ ,
\eeqn
where $u_1(v)$ and $u_2(v)$ are the values of $u$ at which a given 
$v$=constant line intercepts the sphere $r=R$, the surface of the star,  
with $u_2>u_1$ (see Fig.~\ref{fig:u1u2}). 
\vskip 1cm

The MHD equations, in terms of $\hat{u}$ and $\hat{v}$, take the form
\beqn
  \partial_{\hat{t}} \hat{B} &=& -\frac{2\alpha \lambda r^2 (3\cos^2 \theta +1)}
{u_2(v)-u_1(v)} \partial_{\hat{u}} \Omega \label{eq:dtBl2} \\
  \partial_{\hat{t}} \hat{\Omega} &=& -\frac{2 \alpha^2 \lambda}{\alpha_R}
\left[ \frac{r^2 (3\cos^2 \theta+1)}{u_2(v)-u_1(v)} \partial_{\hat{u}} \hat{B}
+ \hat{B} \right] \cr & & \cr
& &  + (\partial_{\hat{t}} \hat{\Omega})_{\rm{vis}} \ .
\label{eq:dtWl2}
\eeqn
Note that the
relativistic factors $\alpha$ and $\lambda$ are functions of $r$. The
variables $r$ and $\theta$ are
regarded as implicit functions of $\hat{u}$ and $\hat{v}$. The
transformation between these two sets of variables are derived
in Appendix~{\ref{app:uv}}.
we only consider the solution with equatorial symmetry. The computation 
domain is $\hat{u} \in [0,1]$ and $\hat{v} \in [0,1]$. The boundary of 
the star is located at $\hat{u}=0$ and $\hat{u}=1$. As in the Newtonian 
case, lines of constant $\hat{v}$ coincide with poloidal field lines.

It is not easy to handle the viscosity term $(\partial_{\hat{t}} 
\hat{\Omega})_{\rm vis}$ numerically in this $(\hat{u},\hat{v})$ 
coordinate system because of the coordinate singularity at 
$\hat{v}=0$ and $\hat{v}=1$. Besides, these particular coordinates 
lose their advantage in the presence of viscosity. Hence
we use the cylindrical-like coordinates 
introduced in Section~\ref{sec:ndl=1} when we study the effect 
of viscosity.

In the absence of viscosity, the energy $\hat{E}$ and angular 
momentum $\hat{J}$ are given by 
\beqn
  \hat{E} &=& \frac{1}{2\sqrt{3}} \int_0^1 d\hat{v} \frac{u_2(v)-u_1(v)}
{R^2} \hat{\epsilon}_2 (\hat{v}) \\
  \hat{J} &=& \frac{\alpha_R}{\sqrt{3}} \int_0^1 d\hat{v} \frac{u_2(v)-
u_1(v)}{R^2} \hat{j}_2 (\hat{v}) \ ,
\eeqn
where 
\beqn
  \hat{\epsilon}_2 (\hat{v}) &=& \hat{\epsilon}_2^{\rm rot} (\hat{v}) +
\hat{\epsilon}_2^{\rm mag} (\hat{v}) \ , \label{eq:e2v} \\ 
  \hat{\epsilon}_2^{\rm rot} (\hat{v}) &=& \int_0^1 d\hat{u} \frac{\alpha_R 
\sin^2 \theta}{\alpha^2 \lambda (3\cos^2 \theta + 1)} \hat{\Omega}^2 \ ,
\label{eq:e2rot} \\
  \hat{\epsilon}_2^{\rm mag} (\hat{v}) &=& \int_0^1 d\hat{u} \frac{\sin^2 \theta}
{\alpha \lambda (3\cos^2 \theta + 1)} \hat{B}^2 \ , \label{eq:e2mag} \\
  \hat{j}_2 (\hat{v}) &=& \int_0^1 d\hat{u} \frac{\Omega \sin^2 \theta}
{\alpha^2 \lambda (3\cos^2 \theta + 1)} \ . \label{eq:j2v}
\eeqn
It follows from Eqs.~(\ref{eq:dtBl2}) and~(\ref{eq:dtWl2}) that 
in the absence of viscosity, the functions $\hat{\epsilon}_2 (\hat{v})$ 
and $\hat{j}_2 (\hat{v})$ are conserved for each $\hat{v}$.

\section{Numerical Results}
\label{sec:numerics}

We numerically integrate the two sets of differential 
equations~(\ref{eq:dtBl1}),~(\ref{eq:dtWl1}) and~(\ref{eq:dtBl2}),~(\ref{eq:dtWl2}) 
by the iterated Crank-Nicholson method~\cite{teukolsky00}. A more detailed 
description of our finite differencing scheme is given in Appendix~\ref{app:FD}. 
In this section, we report our numerical results. Table~\ref{tab:cases} summaries 
the cases we have studied.

\begin{table}
\caption{Summary of the cases we have studied.}
\begin{tabular}{|c|c|c|c|c|}
\hline
 Case & Gravitation & $M/R$ & Poloidal field & Viscosity $\hat{\nu}$  \\
\hline
  Ia   & Newtonian   & $\ll 1$ & $l=1$  & 0  \\
  Ib   & GR & 0.3 & $l=1$ & 0  \\
  Ic   & GR & 0.44 & $l=1$ & 0  \\
  IIa  & Newtonian & $\ll 1$ & $l=2$ & 0  \\
  IIb  & GR & 0.3 & $l=2$ & 0  \\
  IIc  & GR & 0.44 & $l=2$ & 0  \\
  III  & GR & 0.3 & $l=2$ & $0.002$ \\
\hline
\end{tabular}
\label{tab:cases}
\end{table}

\subsection{Without Viscosity (Cases Ia--IIc)}

In the absence of viscosity, we use the special coordinates discussed in 
Sections~\ref{sec:ndl=1} and~\ref{sec:ndl=2} to integrate the nondimensional 
MHD equations.

\subsubsection{$l=1$ initial field (Cases Ia--Ic)} 

As discussed in the Newtonian case, there is no magnetic braking if the initial angular 
velocity distribution is constant on cylinders for the $l=1$ 
initial field. The same result applies for the relativistic case, as is indicated 
from Eq.~(\ref{eq:relB5}).
To study magnetic braking,
we use the following {\it ad hoc} initial differential angular 
velocity profile\footnote{According to Eqs.~(142) and~(\ref{eq:IO}), 
$\Omega_c = 2\, \Omega_{\rm central}$. Also, the specific angular momentum 
increases with $\hat{\varpi}$ for this rotation law. Hence it satisfies 
the Rayleigh criterion of dynamical stability against axisymmetric perturbations.}
\beq
  \hat{\Omega}(0;\hat{\varpi},\hat{z}) = \frac{1}{1+\hat{\varpi}^2}
- \frac{1-\hat{\varpi}^2}{2}\cos \left[ \frac{\pi \hat{z}}{\sqrt{1-
\hat{\varpi}^2}} \right]  \ ,
\label{eq:IO}
\eeq
which is not constant on cylinders. This corresponds to setting the function 
\beq
  C_n(\varpi) = \left \{ \begin{array}{ll} 
   \frac{\Omega_c R}{2v_A} \left( 1- \frac{\varpi^2}{R^2} \right) & \ \ n=1 \\
   0 & \ \ n>1 \end{array} \right. 
\eeq
in Eq.~(\ref{eq:Cn}). Hence in the Newtonian limit ($M/R \ll 1$), 
the toroidal field and angular velocity oscillate with the fundamental 
eigenfrequency on each cylindrical surface, given by Eq.~(\ref{eq:omegan}) 
for $n=1$. The analytic solution in this limit is 
\beqn
  \hat{\Omega}(\hat{t};\hat{\varpi},\hat{z}) &=& \frac{1}{1+\hat{\varpi}^2}
- \frac{1-\hat{\varpi}^2}{2}\cos \left( \frac{\pi \hat{z}}{\sqrt{1-
\hat{\varpi}^2}} \right) \times \cr
& &  \cos \left( \frac{\pi \hat{t}}{\sqrt{1-
\hat{\varpi}^2}} \right) \ , \label{eq:anaOm} \\
  \hat{B}(\hat{t};\hat{\varpi},\hat{z}) &=& \frac{1-\hat{\varpi}^2}{2} 
\sin \left( \frac{\pi \hat{z}}{\sqrt{1-\hat{\varpi}^2}} \right) \times \cr
& & \sin \left( \frac{\pi \hat{t}}{\sqrt{1-\hat{\varpi}^2}} \right) \ .
\label{eq:anaB}
\eeqn
It follows from Eqs.~(\ref{eq:e1rot})--(\ref{eq:jl1}) that 
\beqn
  \hat{\epsilon}_1^{\rm rot}(\hat{t};\hat{\varpi}) &=& \frac{1}{(1+\hat{\varpi}^2)^2}
+\frac{(1-\hat{\varpi}^2)^2}{8} \times \cr
& &  \cos^2 \left( \frac{\pi \hat{t}}
{\sqrt{1-\hat{\varpi}^2}} \right) \ , \label{eq:anae1rot} \\
  \hat{\epsilon}_1^{\rm mag}(\hat{t};\hat{\varpi}) &=& \frac{(1-\hat{\varpi}^2)^2}
{8} \sin^2 \left( \frac{\pi \hat{t}} {\sqrt{1-\hat{\varpi}^2}} \right) \ , \\
  \hat{\epsilon}_1(\hat{\varpi}) &=& \frac{1}{(1+\hat{\varpi}^2)^2}
+\frac{(1-\hat{\varpi}^2)^2}{8} \ , \label{eq:anae1mag} \\
  \hat{j}_1(\hat{\varpi}) &=& \frac{1}{1+\hat{\varpi}^2} \ . 
\eeqn

\begin{figure}
\vskip 1cm
\includegraphics[width=8cm]{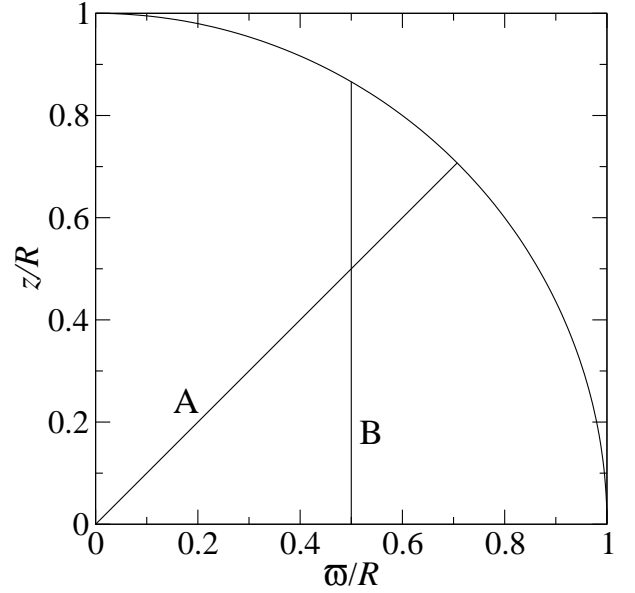}
\vskip 0.5cm
\caption{The $\theta=45^{\circ}$ line (line A) and $\varpi/R=0.5$ line
(line B). The circular arc is the surface of the star $r=R$.}
\label{fig:lines}
\end{figure}

\begin{figure}
\vskip 1cm
\includegraphics[width=8cm]{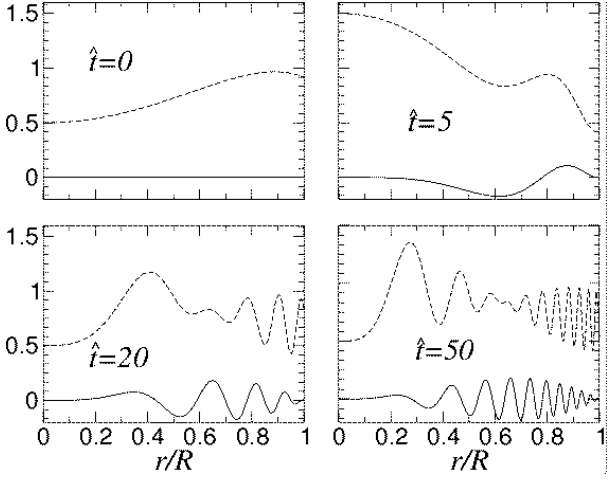}
\vskip 0.5cm
\caption{Snapshots of $\bar{B}$ (solid lines) and
$\hat{\Omega}$ (dashed lines) along the $\theta=45^{\circ}$ line
(line A of Fig.~\ref{fig:lines}) for a Newtonian
star ($M/R \ll 1$). The initial magnetic field is the $l=1$ field given
by Eq.~(\ref{eq:IMFrel}). As expected, small-scale structures build up at
late times.}
\label{fig:analsol1}
\end{figure}

\begin{figure}
\vskip 1cm
\includegraphics[width=8cm]{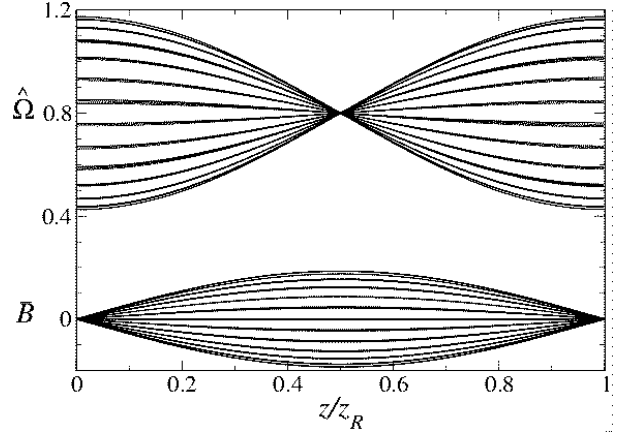}
\vskip 0.5cm
\caption{Snapshots of $\bar{B}$ (lower curves) and
$\hat{\Omega}$ (upper curves) along the $\varpi/R=0.5$ line (line B in 
Fig.~\ref{fig:lines}) for a
Newtonian star ($M/R \ll 1$). The surface of the star is located
at $z=z_R=\sqrt{R^2-\varpi^2}=R/\sqrt{2}$.
Each curve represents the profile
at a particular time. Both $\bar{B}$ and $\hat{\Omega}$ oscillate
sinusoidally with a period $\hat{T}=\sqrt{3}=1.73$.
Curves are plotted at time intervals $\Delta \hat{t}=0.2$ from $\hat{t}=0$ to
$\hat{t}=8.6$ (about 5 oscillation periods).}
\label{fig:analsol2}
\end{figure}

\begin{figure}
\vskip 1cm
\includegraphics[width=8cm]{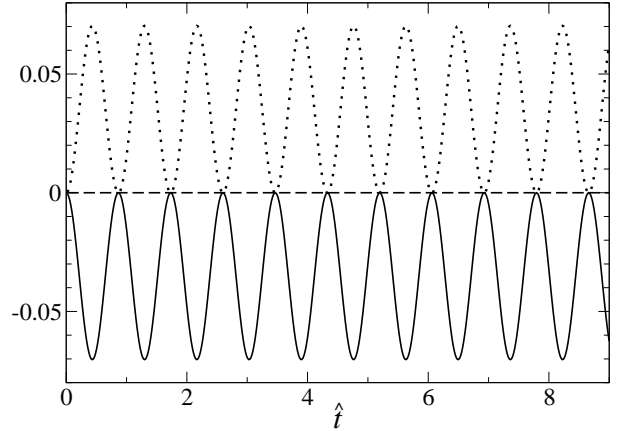}
\vskip 0.5cm
\caption{Energy evolution for a differentially rotating Newtonian star with 
an $l=1$ poloidal magnetic field. The solid line shows 
$\hat{\epsilon}_1^{\rm rot}(\hat{t})-\hat{\epsilon}_1^{\rm rot}(0)$, the dotted line 
shows $\hat{\epsilon}_1^{\rm mag}(\hat{t})$. The dashed line is the sum of the 
two lines, which is $\hat{\epsilon}_1-\hat{\epsilon}_1^{\rm rot}(0)=0$ 
All the reduced energies are evaluated at 
$\hat{\varpi}=0.5$.}
\label{fig:engL1m00}
\end{figure}

We plot the snapshots of $\bar{B}$ and $\hat{\Omega}$ 
as a function of time along two arbitrary lines as 
shown in Fig.~\ref{fig:lines}.
Figure~\ref{fig:analsol1} shows the snapshots of $\bar{B}$ and 
$\hat{\Omega}$ as a function of time along the $\theta=45^{\circ}$ line 
based on the analytic solution given by Eqs.~(\ref{eq:anaOm}) 
and~(\ref{eq:anaB}). We see that small-scale structures develop 
at late times, as predicted from our previous analytic study. 
Figure~\ref{fig:analsol2} shows the snapshots along the $\varpi/R=0.5$ 
line. We see simple sinusoidal oscillations with a period 
$\hat{T}=\sqrt{3}=1.73$. No small structure is observed since 
we are looking in the direction along a particular poloidal field 
line. 

Since $\hat{j}_1(\hat{\varpi})$ is conserved, we can define 
a mean angular velocity for each value of $\hat{\varpi}$ by 
\beq
   \hat{\Omega}_{\rm mean} (\hat{\varpi}) = \hat{j}_1(\hat{\varpi})/
\hat{I}_1(\hat{\varpi}) \ , 
\label{eq:meanOm}
\eeq
where 
\beq
  \hat{I}_1(\varpi) = \int_0^1 \frac{d \hat{u}}{\alpha^2 \lambda} \ .
\eeq
In our Newtonian example, we have 
\beq
  \hat{\Omega}_{\rm mean} (\hat{\varpi}) = \frac{1}{1+\hat{\varpi}^2} \ .
\eeq
Equation~(\ref{eq:anaOm}) indicates that for each $\hat{\varpi}$, the angular 
velocity $\hat{\Omega}$ oscillates about $\hat{\Omega}_{\rm mean}$ with a 
period $\hat{T}=2 \sqrt{1-\hat{\varpi}^2}$. The first term of 
Eq.~(\ref{eq:anae1rot}) is the reduced rotational kinetic energy associated 
with $\hat{\Omega}_{\rm mean}$, which is independent of time. The 
kinetic energy associated with the differential rotation transfers 
back and froth with the energy associated with the toroidal magnetic 
field with a period $\hat{T}/2$ (see Fig.~\ref{fig:engL1m00}).

There is no analytic solution to the MHD equations when the star 
is relativistic. We solve the MHD 
equations~(\ref{eq:dtBl1}) and~(\ref{eq:dtBl2}) numerically in this case. 
The appearance of small-scale structures does not cause any problem 
in our numerical code, because we have chosen suitable coordinates 
to avoid taking derivatives perpendicular to the poloidal field lines. 
To check our numerical integrations, we perform second-order convergence tests and, 
in addition, monitor the conserved integrals 
$\hat{\epsilon}_1(\hat{\varpi})$ and $\hat{j}_1(\hat{\varpi})$ defined 
in Eqs.~(\ref{eq:el1}) and~(\ref{eq:jl1}). We find that 
the fractional change of $\hat{\epsilon}_1$ to be less than $10^{-4}$ 
from time $\hat{t}=0$ to $\hat{t}=100$ for various values of the compaction 
ratio $M/R$. The fractional change of $\hat{j}_1$ is less than $10^{-5}$ over 
the same period of time.

\begin{figure}
\vskip 1cm
\includegraphics[width=8cm]{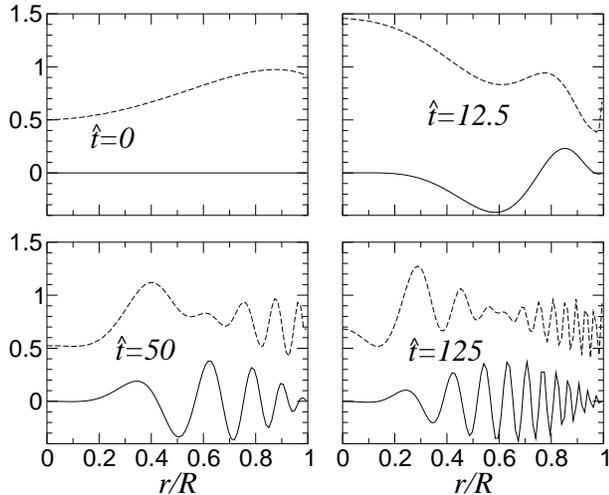}
\vskip 0.5cm
\caption{Same as Fig.~\ref{fig:analsol1} but for a relativistic star with 
$M/R=0.3$. The behavior is qualitatively the same as the Newtonian case, but the
amplitude of oscillation is larger.}
\label{fig:L1snap1m03}
\end{figure}

\begin{figure}
\vskip 1cm
\includegraphics[width=8cm]{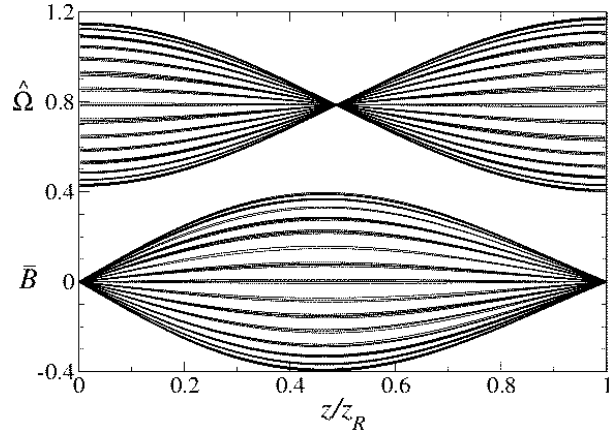}
\vskip 0.5cm
\caption{Snapshots of $\bar{B}$ (lower curves) and
$\hat{\Omega}$ (upper curves) along the $\varpi/R=0.5$ line for
a relativistic star with $M/R=0.3$. As in the Newtonian case,
both $\bar{B}$ and $\hat{\Omega}$ oscillate periodically. The oscillation
period is $\hat{T} = 4.3$, which is longer than the Newtonian case because 
of the relativistic time dilation (see the text). Curves are plotted
at time intervals $\Delta \hat{t}=0.4$ from $\hat{t}=0$ to $\hat{t}=21.2$
(about 5 oscillation periods).}
\label{fig:L1snap2m03}
\end{figure}

\begin{figure}
\vskip 1cm
\includegraphics[width=8cm]{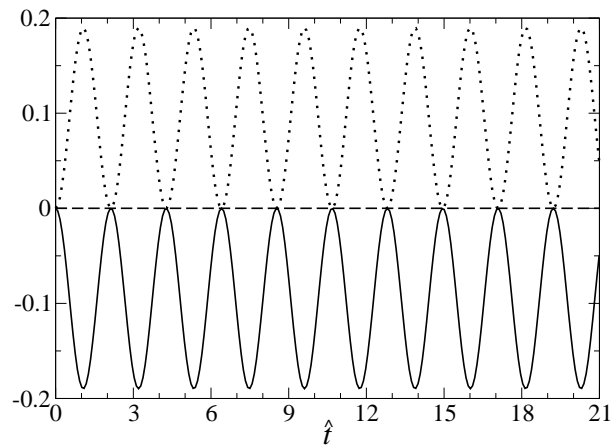}
\vskip 0.5cm
\caption{Same as Fig.~\ref{fig:engL1m00} but for a relativistic star with 
$M/R=0.3$.}
\label{fig:engL1m03}
\end{figure}

Figure~\ref{fig:L1snap1m03} shows the snapshots of $\bar{B}$ and $\hat{\Omega}$ 
for a relativistic star with $M/R=0.3$ along the $\theta=45^{\circ}$ line. We see 
that the behavior is qualitatively the same as the Newtonian case, but the 
amplitude of the toroidal field is larger by a factor of 2. 
Figure~\ref{fig:L1snap2m03} shows the
snapshots along the $\varpi/R=0.5$ coordinate line. The 
initial angular velocity distribution [Eq.~(\ref{eq:IO})] no longer corresponds
to the fundamental mode of the relativistic equations, but it is still very 
close. The oscillation frequency 
is found to be $\hat{T}=4.3$, which is longer than the Newtonian period 
by a factor of 2.5. The corresponding reduced energy is similar to 
Fig.~\ref{fig:engL1m00}, but the amplitudes are larger and 
the period is longer by a factor of 2.5 (Fig.~\ref{fig:engL1m03}).
This can be understood by rewriting Eqs.~(\ref{eq:relB5}) 
and~(\ref{eq:relMHD5}) as 
\beq
  \partial^2_{\hat{t}} \hat{B} = \frac{\alpha^3 \lambda^2}{\alpha_R} 
\left[ \partial^2_{\hat{z}} \hat{B} + \frac{\partial_{\hat{z}} (\alpha^2 \lambda)}
{\alpha^2 \lambda} \hat{B} \right] \ .
\eeq
If the background metric varies slowly along the field lines so that the term 
$\partial_{\hat{z}} (\alpha^2 \lambda)/\alpha^2 \lambda \ll 1$, the equation 
simplifies to 
\beq
  \partial_{\hat \tau}^2 \hat{B} \approx \partial_{\hat{z}}^2 \hat{B} \ ,
\label{eq:scalingeq}
\eeq
where 
\beq
  \hat{\tau} = \left( \frac{\alpha^3 \lambda^2}{\alpha_R} \right)^{1/2} \hat{t} \ .
\label{eq:tscaling}
\eeq
Note that Eq.~(\ref{eq:scalingeq}) does not contain explicitly the relativistic 
factors $\alpha$ and $\lambda$, which are absorbed in the new time variable 
$\hat{\tau}$ and $\hat{B}$ [see Eq.~(\ref{eq:Bbar})]. Hence in this approximation, 
the relativistic corrections are that  
(a) time is dilated according to Eq.~(\ref{eq:tscaling}), and (b) the toroidal field 
$B^{\hat{\phi}}$ is increased by a factor of $1/\alpha$ for a given initial 
angular velocity profile $\Omega(0;\varpi,z)$ according to 
Eq.~(\ref{eq:Bbar}). Following this analysis, 
we expect that for $M/R=0.3$, the maximum value of $\bar{B}$ on the 
$\hat{\varpi}=0.5$ line should be about 1.9 times 
larger than the Newtonian value, and the oscillation period to be longer 
by about a factor of 2.5. Our numerical result gives a factor of 2 for the 
amplitude and 
2.5 for the oscillation period on the cylinder $\varpi/R=0.5$. This indicates 
that the background spacetime metric for this star changes slowly along the 
poloidal field lines and so the behavior of magnetic braking roughly agrees 
with the above simple analysis.

\begin{figure}
\vskip 1cm
\includegraphics[width=8cm]{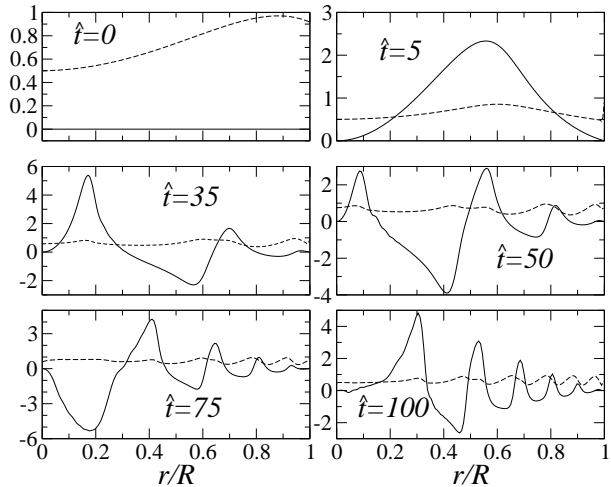}
\vskip 0.5cm
\caption{Same as Fig.~\ref{fig:analsol1} but for a relativistic star
with $M/R=0.44$, approaching
the Buchdahl limit $M/R \rightarrow 4/9$. The behavior of $\bar{B}$
and $\hat{\Omega}$ is qualitative very different from the previous two cases.
However, small-scale structure still appears at late times.}
\label{fig:L1snap1m044}
\end{figure}

\begin{figure}
\vskip 1cm
\includegraphics[width=8cm]{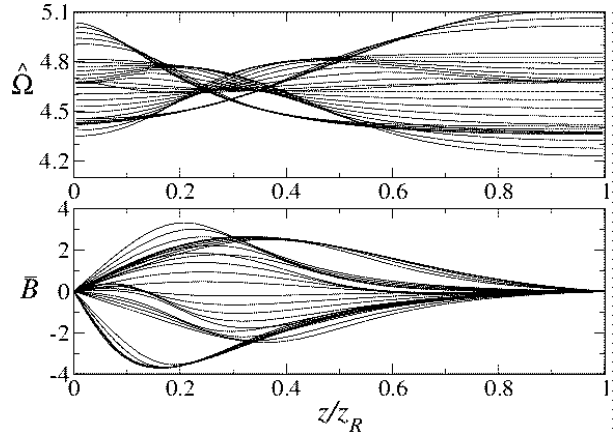}
\vskip 0.5cm
\caption{Snapshots of $\bar{B}$ (lower panel) and
$\hat{\Omega}$ (upper panel) along the $\varpi/R=0.5$ line for
a relativistic star with $M/R=0.44$. Analysis 
with FFT of the time series reveals that the initial data consists of 
mainly three modes with frequencies 
$\hat{f}_1=0.0356$, $\hat{f}_2=0.0622$ and $\hat{f}_3=0.0979$. 
Curves are plotted
at time intervals $\Delta \hat{t}=5$ from $\hat{t}=0$ to 
$\hat{t}=140$ (about 5 oscillation periods of the fundamental mode).}
\label{fig:L1snap2m044}
\end{figure}

\begin{figure}
\vskip 1cm
\includegraphics[width=8cm]{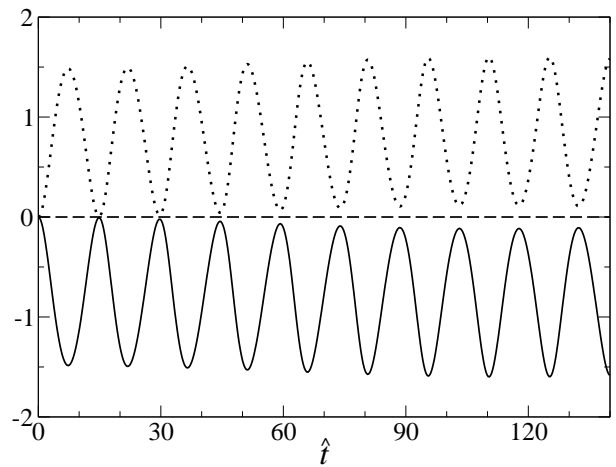}
\vskip 0.5cm
\caption{Same as Fig.~\ref{fig:engL1m00} but for a relativistic star with 
$M/R=0.44$.}
\label{fig:engL1m044}
\end{figure}

To explore strong gravity regime, we consider an extreme relativistic 
star with $M/R=0.44$, approaching the Buchdahl limit $M/R \rightarrow 4/9$. 
Recall that at the Buchdahl limit, the pressure at the center of the star 
($r=0$) becomes infinite, which causes the lapse function $\alpha$ vanish 
at the center. We therefore expect to see a very different 
evolution pattern of $\bar{B}$ and $\hat{\Omega}$. 
Figure~\ref{fig:L1snap1m044} shows the snapshots of $\bar{B}$ and $\hat{\Omega}$ 
along the $\theta=45^{\circ}$ line. We see that although the pattern of the 
curves is very different from the previous two cases, small-scale structures 
still develop at late times. The amplitude of the toroidal field is much larger 
and the oscillation timescale is much longer at small $r$ than that at large 
$r$. It follows from Eqs.~(\ref{eq:relB5}) and~(\ref{eq:relMHD5}) that we 
can define the effective local Alfv\'en speed as 
\beq
  v_{\rm eff} (r) = \sqrt{\frac{\alpha^3 \lambda^2}{\alpha_R}} \left(
\frac{B_0}{\sqrt{4\pi \rho}}\right) = \frac{\hat{\tau}}{\hat{t}} v_A \ .
\label{eq:effva}
\eeq
Hence the Alfv\'en speed is very small at the center when the Buchdahl 
limit is approached, which explains why the oscillation period is long 
close to the center.

Figure~\ref{fig:L1snap2m044} shows the snapshots of $\bar{B}$ and $\hat{\Omega}$
on the $\varpi/R=0.5$ plane. For this extreme relativistic star, the solution 
of the MHD equations with the initial angular velocity distribution~(\ref{eq:IO}) 
is no longer close to the fundamental mode. We perform a fast Fourier transform 
(FFT) on the time series $\bar{B}(\hat{t};0.5,\hat{u})$ for three different values 
of $\hat{u}$. We find that each of the three sets of time series consist 
mainly of three 
frequencies $\hat{f}_1=0.0356$, $\hat{f}_2=0.0622$ and $\hat{f}_3=0.0979$. We 
also see that the peak of the toroidal field shifts with increasing time 
to a lower value of $z$. 
This can be explained by the following physical argument. The initial angular 
velocity can be regarded as a superposition of waves of $B^{\hat{\phi}}$ traveling 
to $z=-z_R=-\sqrt{R^2-\varpi^2}$ and $z=z_R$. The waves are reflected at the two 
points and travel backward. Since the value of $r$ is smaller at smaller $z$, it 
follows from Eq.~(\ref{eq:effva}) that the local Alfv\'en speed decreases as $z$ 
decreases. However, the equatorial symmetry condition on $\Omega$ forces the 
toroidal field to vanish at the equator. 
As a result, the waves pile up near the equator, causing the peak 
to shift to a smaller value of $z$.

From Eq.~(\ref{eq:meanOm}) we found $\hat{\Omega}_{\rm mean}=0.64$. 
Even though the $\hat{\Omega}$ in Fig.~\ref{fig:L1snap2m044} consists of 
several modes, it still oscillates about this mean angular velocity.
As in the previous cases, the rotational kinetic energy associated with the 
differential rotation transfers back and fro to the energy associated 
with the toroidal magnetic field (see Fig.~\ref{fig:engL1m044}).

\subsubsection{$l=2$ initial field (Cases IIa--IIc)}

For the $l=2$ field, we numerically integrate Eqs.~(\ref{eq:dtBl2}) 
and~(\ref{eq:dtWl2}) in the $(\hat{u},\hat{v})$ coordinate system 
introduced in Section~\ref{sec:ndl=2}. The advantage of using these 
rather complicated coordinates is twofold. First, we 
only need to integrate a set of decoupled 1+1 evolution equations. 
Second, more importantly, small-scale structures will develop at late times 
in a direction across the poloidal field lines. Using other coordinate 
systems (e.g.\ standard Cartesian, spherical or cylindrical coordinates) 
will involve taking derivatives across the poloidal field lines, 
which will be numerically inaccurate when the finite-sized grid can 
no longer resolve the growth of small-scale structures. We will demonstrate 
this numerical inaccuracy in Section~\ref{sec:numvis}. To estimate 
the numerical error of our code, we monitor 
the conserved integrals $\hat{\epsilon}_2(\hat{v})$ and 
$\hat{j}_2(\hat{v})$ defined in Eqs.~(\ref{eq:e2v}) and~(\ref{eq:j2v}) 
as well as second order convergence. We find that the 
fractional variation of $\hat{\epsilon}_2$ and $\hat{j}_2$ is 
less than 0.3\% for all of our runs with our canonical grid allocation.

In the literature, the rotation law
\beq
  u^t u_{\phi} = A^2 (\Omega_c - \Omega)
\eeq
is often used to model differentially rotating relativistic stars 
(see, e.g.\ Ref.~\cite{komatsu89}).
Here $A$ is a constant. This law corresponds to, using the
metric~(\ref{bmetric}), the relation
\beq
  \hat{\Omega}=\hat{\Omega}_p=\frac{1}{1+\frac{\hat{r}^2 \sin^2 \theta}
{\alpha^2 \hat{A}^2}} \ ,
\label{eq:rotlaw}
\eeq
where $\hat{A}=A/R$ and we have ignored the frame dragging term 
for simplicity. In 
the Newtonian limit ($\alpha \rightarrow 1$),
$\hat{\Omega}_p$ reduces to the so-called ``$j$-constant'' law~\cite{eriguchi85}. 
Unfortunately, this rotation law does not satisfy the constraint
$B^j(0)\partial_j \Omega=0$ at $r=R$. To ``fix'' this, we choose
a modified rotation profile\footnote{According to Eqs.~(142) 
and~(\ref{eq:IOl2}), $\Omega_c = \Omega_{\rm central}$. This rotation 
law also satisfies the Rayleigh criterion for dynamical stability since 
the specific angular momentum increases with $\hat{\varpi}$.}
\beq
  \hat{\Omega}(0;\hat{r},\theta) = \hat{\Omega}_p(\hat{r},\theta)
f(\hat{r}) + \hat{\Omega}_b(\hat{r},\theta) [1-f(\hat{r})] \ ,
\label{eq:IOl2}
\eeq
where $\hat{\Omega}_b$ is a function that satisfies the constraint
at $r=R$ and $f$ is a function which is close to 1 and drops rapidly
to 0 at $\hat{r}=1$. We also require $f'(1)=0$. Specifically, we choose
\beqn
  f(\hat{r}) &=& \frac{f_F(\hat{r})-f_F(1)+\frac{1}{2} f'_F(1) (1-\hat{r}^2)}
{f_F(0)-f_F(1)+\frac{1}{2}f'_F(1)} \\
  f_F(\hat{r}) &=& \frac{1}{1+\exp [ (\hat{r}-b)/\epsilon]} \ ,
\eeqn
where $b$ and $\epsilon$ are constant parameters which we set to be
$b=0.8$ and $1/\epsilon=25$. We set
\beq
  \hat{\Omega}_b(\hat{r},\theta) = \langle \hat{\Omega}_p(1,\theta) \rangle
\hat{r} (2-\hat{r}) \ ,
\eeq
where
\beqn
  \langle \hat{\Omega}_p(1,\theta) \rangle &=& \frac{1}{4\pi} \int_0^{2\pi}
d\phi \int_0^{\pi} d\theta \sin \theta \hat{\Omega}_p(1,\theta) \\
  &=& \frac{\alpha_R^2 \hat{A}^2}{\sqrt{1+\alpha_R^2 \hat{A}^2}}
\tanh^{-1} \left(\frac{1}{\sqrt{1+\alpha_R^2 \hat{A}^2}}\right) \ \ \ \ \ \ \ 
\eeqn
We set the parameter $\hat{A}=1$ in all of our numerical calculations.
The value of $\hat{\Omega}(0;r,\theta)$ chosen in this way is close 
to $\hat{\Omega}_p(r,\theta)$ except near the surface of the star, 
where the modification makes it satisfy the 
boundary condition $B^j(0) \partial_j \Omega=0$ at $r=R$. 

\begin{figure}
\vskip 1cm
\includegraphics[width=8cm]{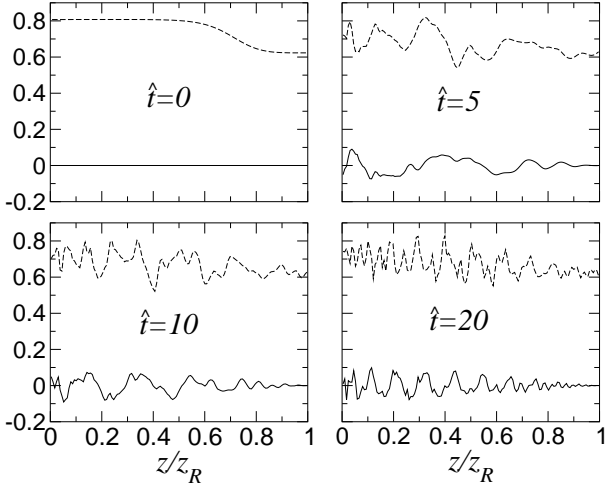}
\vskip 0.5cm
\caption{Snapshots of $\bar{B}$ (lower curves) and
$\hat{\Omega}$ (upper curve) along the $\varpi/R=0.5$ line for 
a Newtonian star ($M/R \ll 1$). The initial magnetic field 
is the $l=2$ field given by Eq.~(\ref{eq:IMFrel2}). The surface of that star 
is located at $z=z_R=\sqrt{R^2-\varpi^2}$. Since the direction does not 
line up with the field lines, the build-up of small-scale structure 
is observed, as expected.}
\label{fig:L2snap1m00}
\end{figure} 

Figure~\ref{fig:L2snap1m00} shows the evolution of the toroidal field 
and angular velocity along the line $\varpi/R=0.5$ for a 
Newtonian star ($M/R \ll 1$). Magnetic braking causes 
the toroidal field and angular velocity to oscillate. We see that 
small-scale structures develop at late times, which is expected since 
the direction does not line up with any poloidal field line, while 
coherent oscillations are restricted to field lines. 

\begin{figure}
\vskip 1cm
\includegraphics[width=8cm]{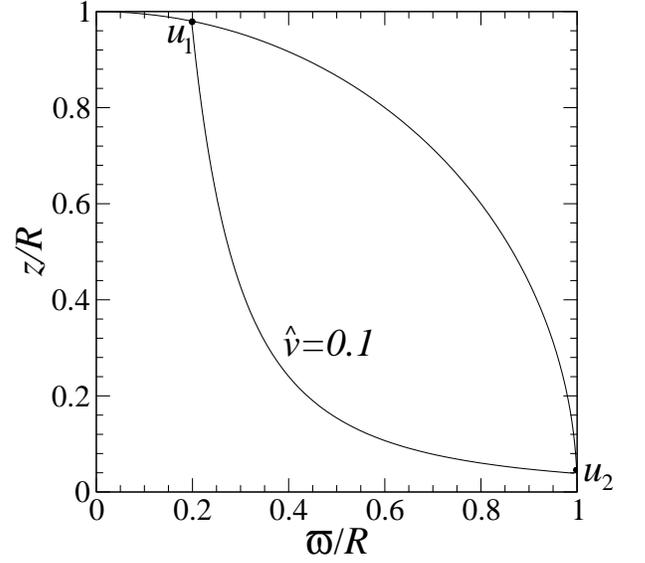}
\vskip 0.5cm
\caption{The $\hat{v}=0.1$ line.}
\label{fig:v01}
\end{figure}

\begin{figure}
\vskip 1cm
\includegraphics[width=8cm]{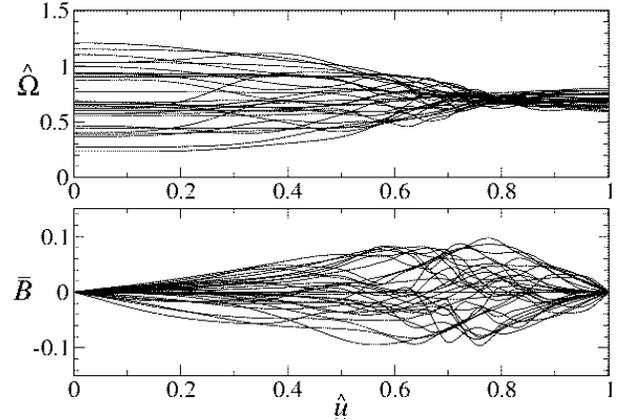}
\vskip 0.5cm
\caption{Snapshots of $\bar{B}$ (lower panel) and
$\hat{\Omega}$ (upper panel) along the $\hat{v}=0.1$ line for
a Newtonian star ($M/R \ll 1$). The FFT spectrum indicates that  
the first four eigenfrequencies to be approximately 
0.35, 0.65, 0.95 and 1.25. Curves are plotted
at time intervals $\Delta \hat{t}=0.5$ from $\hat{t}=0$ to $\hat{t}=15$ 
(about 5 oscillation periods of the fundamental mode).}
\label{fig:L2snap2m00}
\end{figure}

\begin{figure}
\vskip 1cm
\includegraphics[width=8cm]{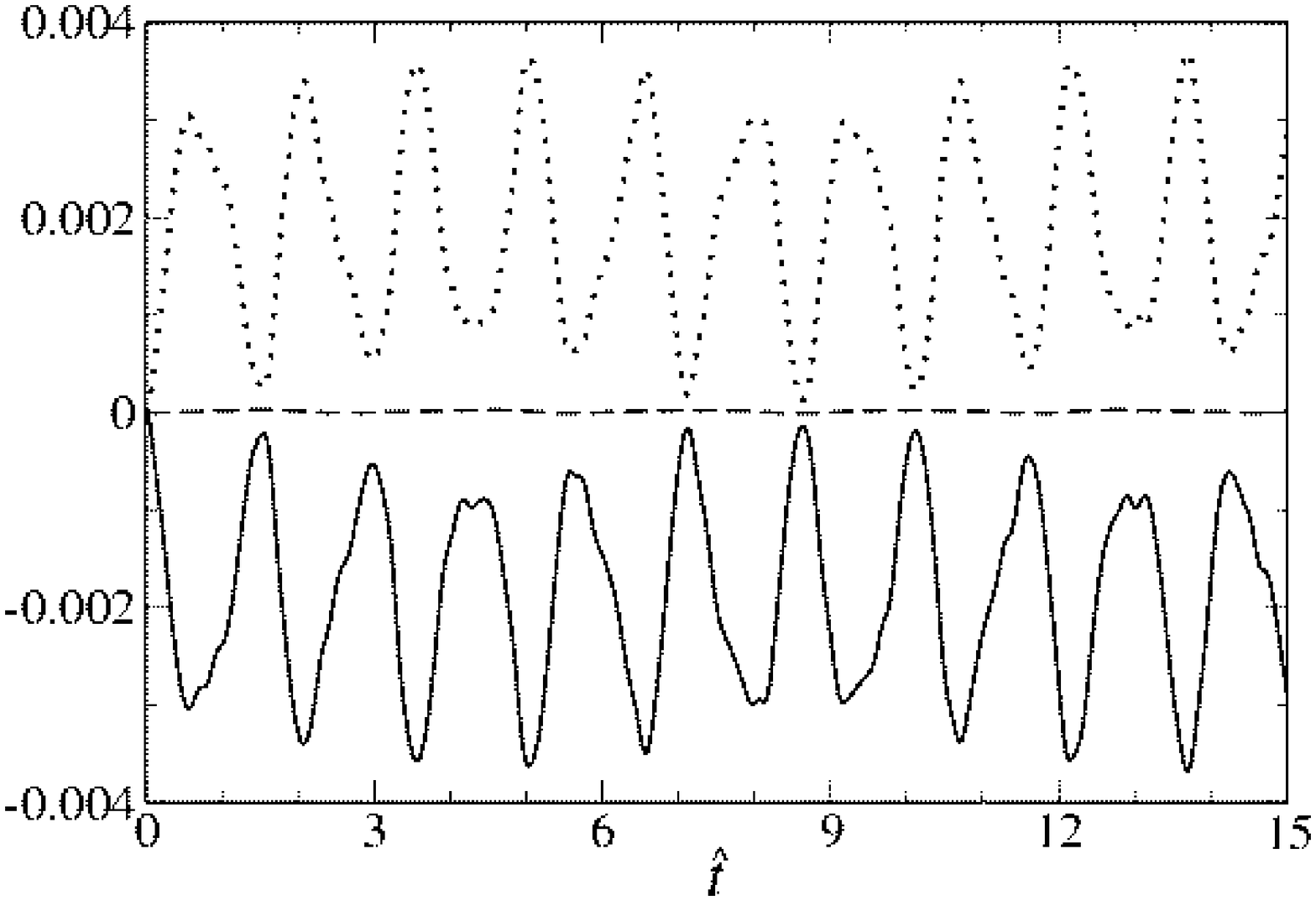}
\vskip 0.5cm
\caption{Evolution of the reduced energies for a Newtonian star with $l=2$ 
initial magnetic field. The solid line is 
$\hat{\epsilon}_2^{\rm rot}(\hat{t})-\hat{\epsilon}_2^{\rm rot}(0)$, the dotted line 
is $\hat{\epsilon}_2^{\rm mag}(\hat{t})$, and the dashed line is 
$\hat{\epsilon}_2-\hat{\epsilon}_2^{\rm rot}(0)$. All the reduced energies 
are evaluated at $\hat{v}=0.1$.}
\label{fig:engL2m00}
\end{figure}

Figure~\ref{fig:L2snap2m00} shows the snapshots of $\bar{B}$ and 
$\hat{\Omega}$ along the line $\hat{v}=0.1$ (see Fig.~\ref{fig:v01}) 
for the same star. Small-scale structures are not seen along this line
since we are looking in the direction of a poloidal field line. We 
perform FFT on the time series of three chosen points on the line. 
We find that the FFT spectra consist of the same frequencies for 
the three points. 
The lowest four eigenfrequencies are determined to be 0.35, 0.65, 
0.95 and 1.25 in our nondimensional units.

Similar to the $l=1$ cases, we can define, from Eq.~(\ref{eq:j2v}), a 
mean angular velocity at each $\hat{v}$ by 
\beq
  \hat{\Omega}_{\rm mean}(\hat{v}) = \hat{j}_2(\hat{v})/\hat{I}_2(\hat{v}) \ ,
\label{eq:Om_mean2}
\eeq
where 
\beq
\hat{I}_2 (\hat{v}) = \int_0^1 d\hat{u} \frac{\sin^2 \theta}
{\alpha^2 \lambda (3\cos^2 \theta + 1)} \ . \label{eq:I2v}
\eeq
For this Newtonian star, we found $\hat{\Omega}_{\rm mean}(0.1)=0.70$. We see from 
Fig.~\ref{fig:L2snap2m00} that $\hat{\Omega}$ oscillates about this mean 
value. Figure~\ref{fig:engL2m00} shows the evolution of the reduced 
energies at $\hat{v}=0.1$. As expected, we see that both the reduced 
rotational kinetic energy
$\hat{\epsilon}_2^{\rm rot}$ and magnetic energy $\hat{\epsilon}_2^{\rm mag}$ 
oscillate, while keeping the total energy $\hat{\epsilon}_2$ conserved.

\begin{figure}
\vskip 1cm
\includegraphics[width=8cm]{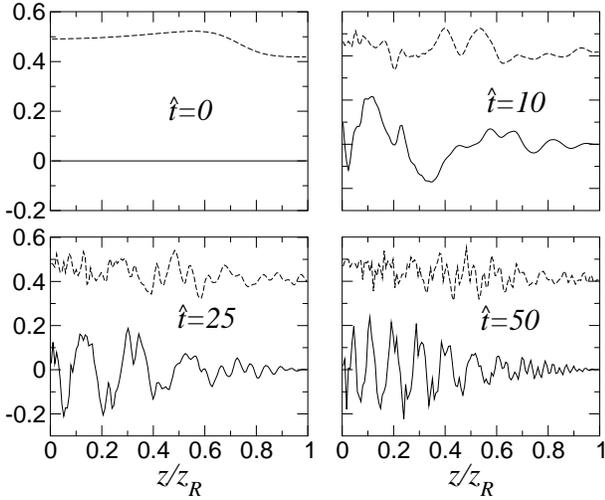}
\vskip 0.5cm
\caption{Same as Fig.~\ref{fig:L2snap1m00} but with $M/R=0.3$.}
\label{fig:L2snap1m03}
\end{figure} 

\begin{figure}
\vskip 1cm
\includegraphics[width=8cm]{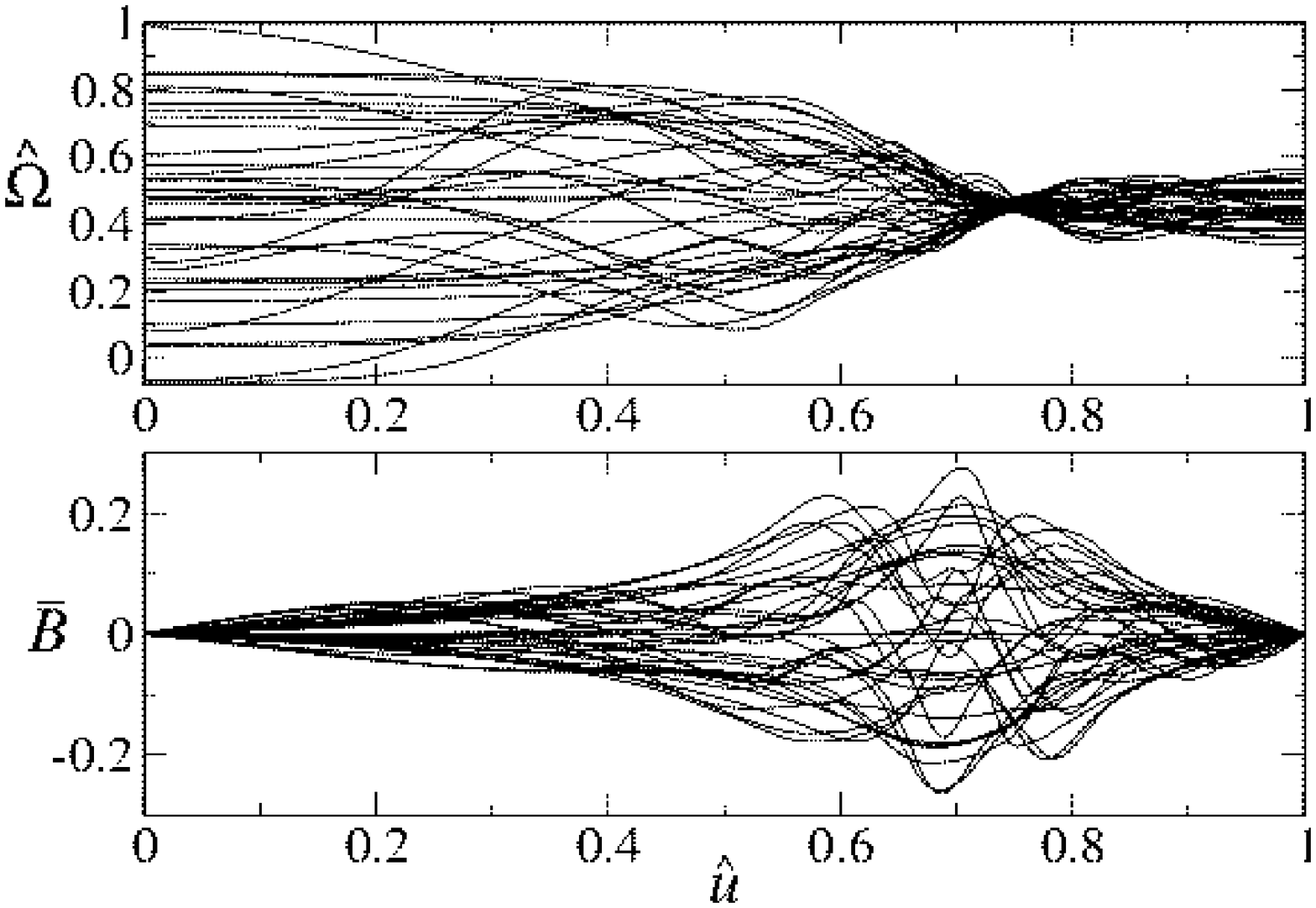}
\vskip 0.5cm
\caption{Snapshots of $\bar{B}$ (lower panel) and
$\hat{\Omega}$ (upper panel) along the $\hat{v}=0.1$ line for
a relativistic star with $M/R=0.3$. The FFT spectrum reveals that
the first four eigenfrequencies to be 0.14, 0.26, 0.38 and 0.62. 
Curves are plotted
at time intervals $\Delta \hat{t}=1$ from $\hat{t}=0$ to $\hat{t}=35$ 
(about 5 oscillation periods of the fundamental mode).}
\label{fig:L2snap2m03}
\end{figure}

\begin{figure}
\vskip 1cm
\includegraphics[width=8cm]{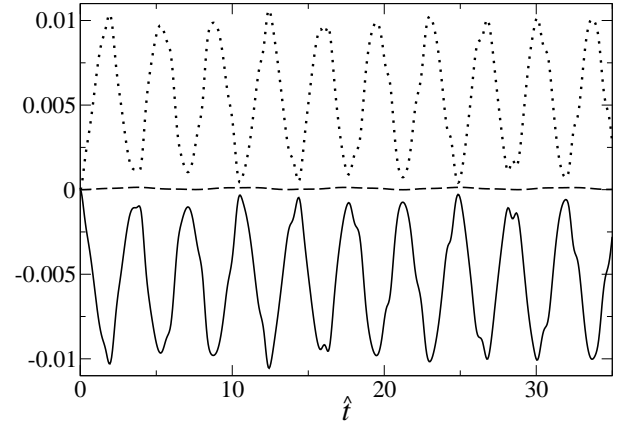}
\vskip 0.5cm
\caption{Same as Fig.~\ref{fig:engL2m00} but for a relativistic star 
with $M/R=0.3$.}
\label{fig:engL2m03}
\end{figure}

Figure~\ref{fig:L2snap1m03} shows the snapshots of $\bar{B}$ and 
$\hat{\Omega}$ along the $\varpi/R=0.5$ line for a relativistic star 
with $M/R=0.3$. The graphs look qualitatively the same as the 
Newtonian situation. 
Similar to the $l=1$ case, it is easy to see 
from Eqs.~(\ref{eq:dtBl2}) 
and~(\ref{eq:dtWl2}) that if the relativistic factors $\alpha$ and 
$\lambda$ do not change significantly inside the star, 
the main relativistic effects are to cause time to dilate according to 
Eq.~(\ref{eq:tscaling}) 
and to increase the amplitude of the toroidal field 
by a factor of $1/\alpha$ for a given set of initial data. 
This explains the similarity of the features 
in Figs.~\ref{fig:L2snap1m00} and~\ref{fig:L2snap1m03}, and that 
small-scale structures appear later for the relativistic star.
Figure~\ref{fig:L2snap2m03} shows the snapshots 
along the $\hat{v}=0.1$ line, and Fig.~\ref{fig:engL2m03} shows the evolution 
of the reduced energies. The angular velocity function $\hat{\Omega}(t;\hat{u},0.1)$ 
oscillates about the mean value $\hat{\Omega}_{\rm mean}(0.1)=0.45$. 
The FFT spectrum reveals the lowest four 
eigenfrequencies to be 0.14, 0.26, 0.38 and 0.62. Note that these 
frequencies are 40\% of those of the Newtonian stars, which agrees  
with the time dilation effect predicted by Eq.~(\ref{eq:tscaling}). However, 
the ratio of the amplitudes of the toroidal field is much larger 
than the predicted value $\alpha$. This is because the two stars 
do not have the same initial angular velocity profile [see 
Eqs.~(\ref{eq:rotlaw}) and~(\ref{eq:IOl2})].

\begin{figure}
\vskip 1cm
\includegraphics[width=8cm]{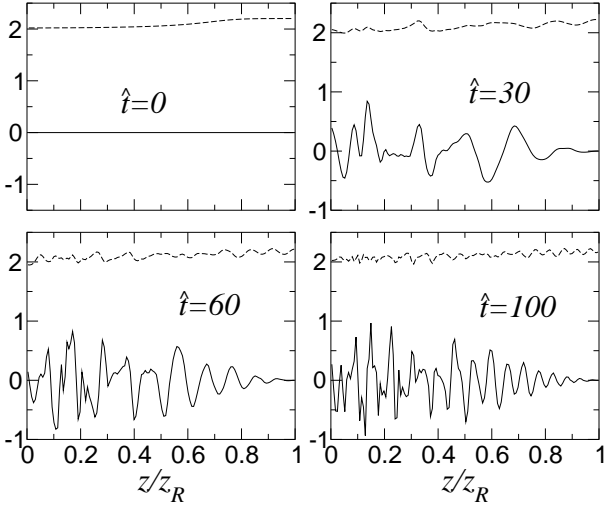}
\vskip 0.5cm
\caption{Same as Fig.~\ref{fig:L2snap1m00} but with $M/R=0.44$ (The upper 
curve is $\hat{\Omega}+2$).}
\label{fig:L2snap1m044}
\end{figure}

\begin{figure}
\vskip 1cm
\includegraphics[width=8cm]{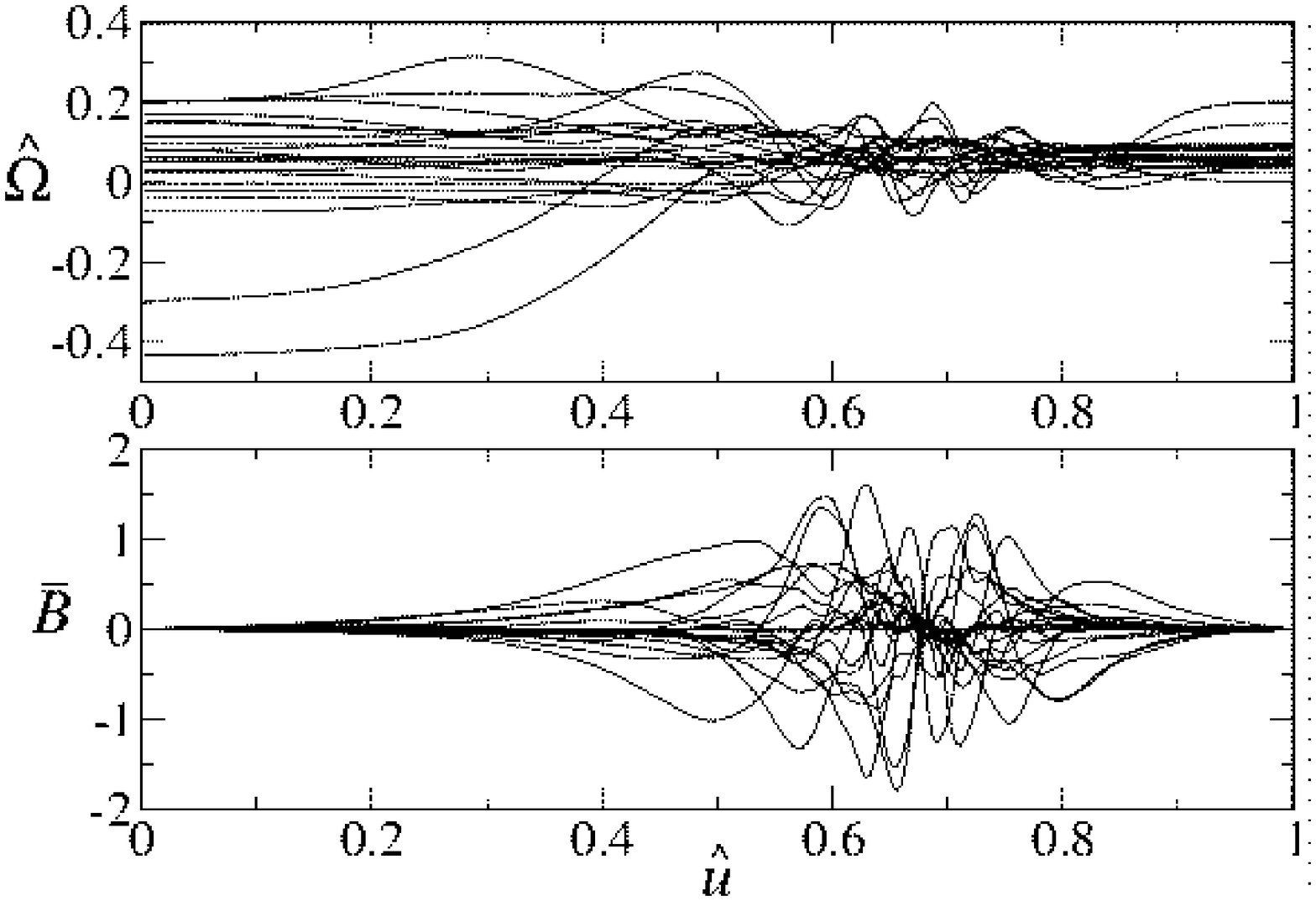}
\vskip 0.5cm
\caption{Snapshots of $\bar{B}$ (lower panel) and
$\hat{\Omega}$ (upper panel) along the $\hat{v}=0.1$ line for
a relativistic star with $M/R=0.44$. The FFT spectrum reveals that
the first four eigenfrequencies to be 0.016, 0.031, 0.045 and 
0.060.  Curves are plotted
at time intervals $\Delta \hat{t}=15$ from $\hat{t}=0$ to $\hat{t}=300$ 
(about 5 oscillation periods of the fundamental mode).}
\label{fig:L2snap2m044}
\end{figure}

\begin{figure}
\vskip 1cm
\includegraphics[width=8cm]{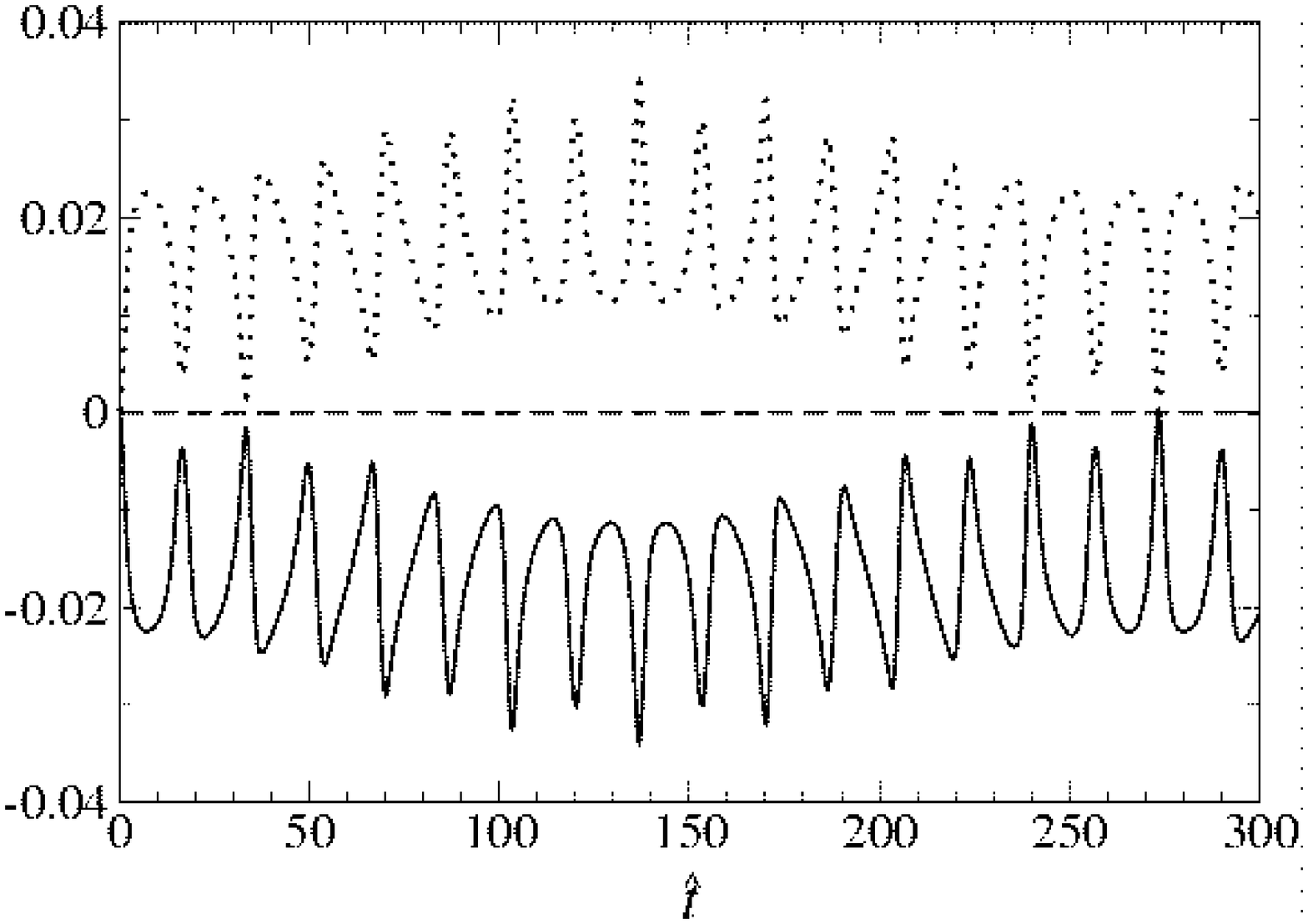}
\vskip 0.5cm
\caption{Same as Fig.~\ref{fig:engL2m00} but for a relativistic star 
with $M/R=0.44$.}
\label{fig:engL2m044}
\end{figure}

Finally, we explore effect of strong gravity by considering a relativistic 
star with $M/R=0.44$, approaching the Buchdahl limit. 
Figures~\ref{fig:L2snap1m044} (\ref{fig:L2snap2m044}) shows the snapshots 
of $\bar{B}$ and $\hat{\Omega}$ along the $\varpi/R=0.5$ ($\hat{v}=0.1$) line. 
Figure~\ref{fig:engL2m044} shows the evolution of the reduced energies 
at $\hat{v}=0.1$. 
We see that the behavior is qualitatively very different from the previous 
two stars. Small-scale structures still develop, but it happens at much 
later time. The oscillation amplitudes of $\bar{B}$ and $\hat{\Omega}$ are 
much larger along the $\hat{v}$=constant lines. 

\begin{figure}
\vskip 1cm
\includegraphics[width=8cm]{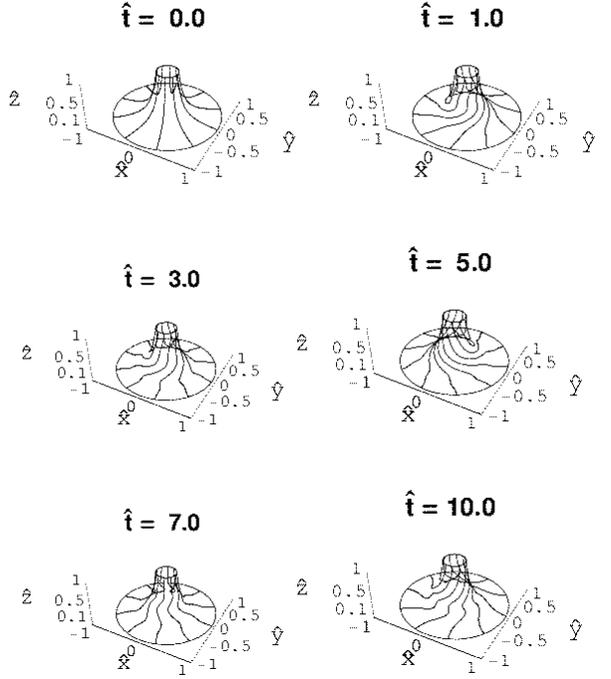}
\vskip 0.5cm
\caption{The magnetic field lines as functions of time along the 
axisymmetric surface $\hat{v}=0.1$ 
for a relativistic star with $M/R=0.3$. 
The poloidal field lines are the same as the field lines at $\hat{t}=0$.
The toroidal field has been multiplied by an arbitrary factor of 
$10 v_A/\Omega_c R$ so that they are more visible. The circles on the upper 
and lower ends are the boundary of the star, i.e., the lines of revolution 
generated by the points $u_1$ and $u_2$ in Fig.~\ref{fig:v01}.}
\label{fig:Blines2}
\end{figure}

The growth of the toroidal field twists the poloidal field lines in the azimuthal 
direction. The magnetic field lines are therefore still confined to the surface 
spanned by the poloidal field lines. 
However, since we consider the situation in which the rotational 
timescale is much longer than the Alfv\'en timescale, the induced toroidal 
field is very small compared to the poloidal field. 
In order to see the small 
twist of the field lines, we multiply the toroidal field $B^{\hat{\phi}}$ by 
an arbitrary factor of $10 v_A/\Omega_c R$ in Fig.~\ref{fig:Blines2}, 
where the field lines on the $\hat{v}=0.1$ 
surface is plotted for a relativistic star with $M/R=0.3$. As expected, 
the oscillations of the toroidal field 
lines cause the total field lines to twist back and forth in the azimuthal 
direction. The field lines of relativistic stars with other compaction 
factors $M/R$ behave in a similar way.

\subsection{Role of Viscosity (Case III)}
\label{sec:numvis}

Our previous results show that the phase mixing arising from magnetic 
braking is  
likely to induce irregular angular velocity flow. A significant 
meridional current will build up eventually, and thus may induce other 
magnetic instabilities and cause turbulence. Magnetic diffusion 
and viscosity will play a significant role on the subsequent evolution. 
It has been shown in Newtonian MHD that magnetic diffusion will 
damp out the angular velocity oscillations along a poloidal line. 
The final state of the star will have a constant angular velocity profile 
on each magnetic surface (see~\cite{spruit99} and references therein). 
Here we wish to discuss the role of viscosity in damping differential 
motion. Microscopic viscosity in a typical neutron star is very 
small, and its timescale is very long compared to the rotational 
period~\cite{cutler87}. Here we may regard 
viscosity as a ``turbulent viscosity,'' which models 
the effects of turbulence via an effective shear viscosity. 
The typical turbulent viscosity is $\nu \sim l \Delta v$ 
(see Ref.~\cite{landau87}, \S~33), where 
$l$ is the length scale of the largest turbulent eddies and $\Delta v$ 
is the velocity fluctuation over the distance $l$.
This turbulent viscosity is much larger than the microscopic 
viscosity, and can affect the flow on a dynamical timescale.

We solve the MHD equations~(\ref{eq:dtBl2}) and~(\ref{eq:dtWl2}) 
with viscosity $\hat{\nu} \neq 0$. We only present one case here, as the result 
is qualitatively similar for all the other cases. The initial poloidal 
field is given 
by the $l=2$ field and the $M/R$ of the star is 0.3. The viscosity 
coefficient $\hat{\nu}$ is set to be 0.002. This value is chosen so that 
the toroidal field and angular velocity can go through several 
Alfv\'en oscillations before the viscosity damps the oscillations. 

\begin{figure}
\vskip 1cm
\includegraphics[width=8cm]{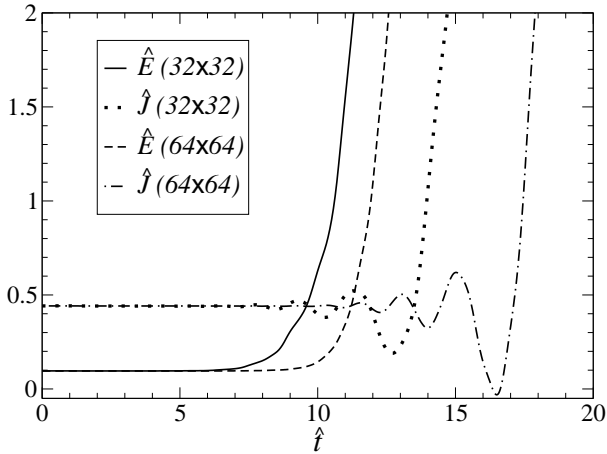}
\vskip 0.5cm
\caption{Energy and angular momentum versus time by evolving 
Eqs.~(\ref{eq:relB2}) and~(\ref{eq:relMHD2}) for the $l=2$ poloidal 
field using the cylindrical-like coordinates $(\hat{\varpi},\hat{u})$ 
in the absence of viscosity. The compaction $M/R$ 
is 0.3. The results for two resolutions 
($32 \times 32$ and $64\times 64$) are plotted. The energy and angular 
momentum derivate from their initial values at late times because 
the finite-sized grid fails to resolve the small-scale structures built 
up by magnetic braking. Increasing the resolution postpones, but does not 
eliminate, the breakdown of conservation.}
\label{fig:intsvis0}
\end{figure}

\begin{figure}
\vskip 1cm
\includegraphics[width=8cm]{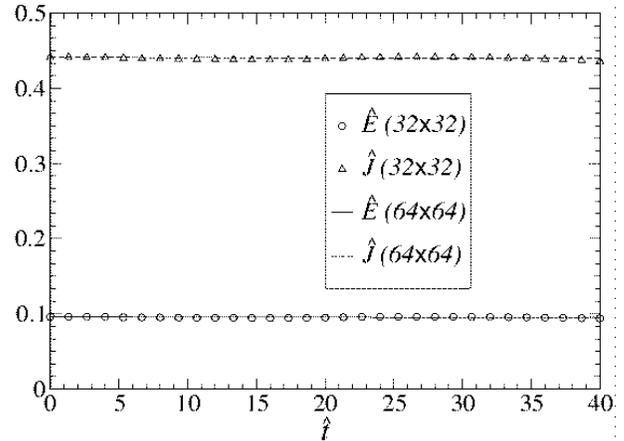}
\vskip 0.5cm
\caption{Same as Fig.~\ref{fig:intsvis0} but with viscosity $\hat{\nu}=0.002$. 
Viscosity suppresses the formation of small-scale structures.
Hence the integrals can be conserved accurately even in 
cylindrical coordinates.}
\label{fig:intsvis}
\end{figure}

As mentioned in Section~\ref{sec:ndl=2}, the $(\hat{u},\hat{v})$ coordinates
are not convenient when viscosity is present. We use the cylindrical-like
coordinates $(\hat{\varpi},\hat{u})$ introduced in Section~\ref{sec:ndl=1}
in this case. In the absence of viscosity, we are not able to integrate
the MHD equations accurately in these coordinates very long because our finite-sized 
grid can no longer resolve the small-scale structures that develop at 
late times. Figure~\ref{fig:intsvis0} 
illustrates the resulting numerical inaccuracy by plotting the energy $\hat{E}$ 
and angular momentum $\hat{J}$ computed from the numerical data 
as a function of time. We evolve the MHD equations with resolutions 
$32\times 32$ and $64 \times 64$. We see that the two conserved integrals 
derivate from their initial values at late times. Doubling the resolution 
does not eliminate, but only postpones, the breakdown of $\hat{E}$ and 
$\hat{J}$ conservation. Hence in the absence 
of viscosity, we must use the $(\hat{u},\hat{v})$ coordinates to ensure 
a stable evolution. However, when we include a small viscosity $\hat{\nu}=0.002$, 
the situation changes drastically. The viscosity suppresses the build up of 
small-scale structure. As a result, the numerical inaccuracy disappears 
(see Fig.~\ref{fig:intsvis}). The fluctuation of energy $\hat{E}$ is 
3\% (0.4\%) in $32\times 32$ ($64\times 64$) resolution. The fluctuation 
of angular momentum $\hat{J}$ is 1\% (0.4\%) in $32\times 32$ ($64\times 64$) 
resolution.

\begin{figure}
\vskip 1cm
\includegraphics[width=8cm]{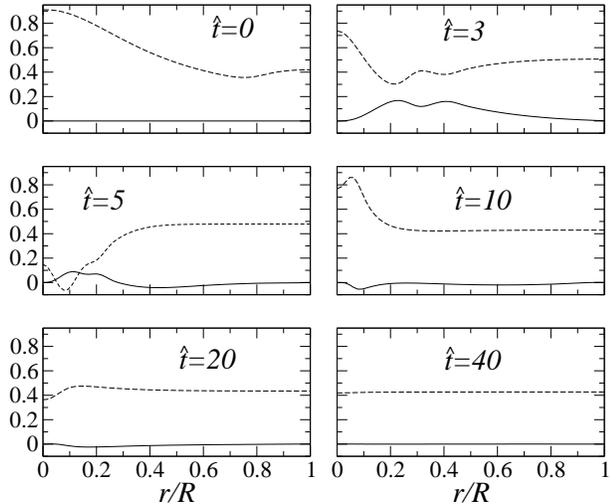}
\vskip 0.5cm
\caption{Snapshots of $\bar{B}$ (solid lines) and
$\hat{\Omega}$ (dashed lines) at the equator in the 
presence of viscosity. The coefficient of viscosity is set to be 
$\hat{\nu} =0.002$. The toroidal field vanishes and the star becomes 
rigidly rotating at late times, as expected.}
\label{fig:L2vissnap}
\end{figure}

\begin{figure}
\vskip 1cm
\includegraphics[width=8cm]{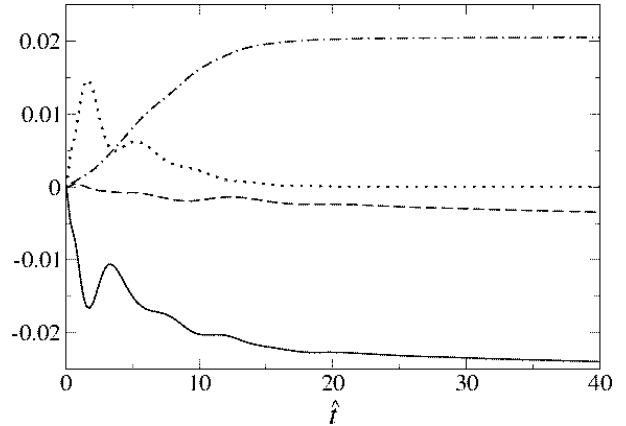}
\caption{Evolution of energies for a relativistic star with $M/R=0.3$ 
in the presence of viscosity. The solid line is $\hat{E}_{\rm rot}(\hat{t})
-\hat{E}_{\rm rot}(0)$, the dotted line is $\hat{E}_{\rm mag}(\hat{t})$, 
the dot-dashed line is $\hat{E}_{\rm vis}(\hat{t})$, 
and the dashed line is the sum of the three lines, i.e.\ 
$\hat{E}_{\rm rot}(\hat{t})+\hat{E}_{\rm mag}(\hat{t})
+\hat{E}_{\rm vis}(\hat{t})-\hat{E}_{\rm rot}(0)$, which should remain 
constant and equal to 0. The deviation from 
zero of the dashed curve is due to truncation error of our code, which 
decreases with increasing resolution.}
\label{fig:engL2vis}
\end{figure}

Figure~\ref{fig:L2vissnap} shows the snapshots of $\hat{B}$ and 
$\hat{\Omega}$ at the equator. We see that the star is driven to
uniform rotation with no toroidal field at late times, as expected. 
Since angular momentum is conserved, we can calculate the final 
angular velocity from Eq.~(\ref{eq:nondJ}), which 
is also valid for the $l=2$ field. We obtain
\beqn
  \hat{\Omega}_{\rm final} &=& \hat{J}/\hat{I} \ , \\
  \hat{I} &=& 3 \alpha_R \int_0^1 d\varpi \, \varpi^3 \sqrt{1-\varpi^2} 
  \int_0^1 \frac{d\hat{u}}{\alpha^2 \lambda} \ .
\eeqn
Not surprisingly, our numerical value of $\hat{\Omega}_{\rm final}$ 
shown in Fig.~\ref{fig:L2vissnap} agrees with the above formula, 
$\hat{\Omega}_{\rm final} = 0.425$.

Finally, Fig.~\ref{fig:engL2vis} shows the evolution of 
$\hat{E}_{\rm rot}(\hat{t})$, $\hat{E}_{\rm mag}(\hat{t})$, and 
$\hat{E}_{\rm vis}(\hat{t})$. We see that a portion of rotational 
energy transfers to magnetic energy of the toroidal field. Both energies 
are eventually turned into heat. The total energy is conserved up to 
the truncation error, which decreases with increasing resolution.

\section{Summary and Discussion}
\label{sec:diss}

We have studied the magnetic braking in differentially rotating, 
relativistic stars. The results we found are very different from 
the models studied in Papers~I and~II. The models in Paper~I 
are incompressible and one-dimensional (infinite cylinders), the models 
in Paper~II are compressible and one-dimensional (infinite cylinders), 
while the models in this paper are incompressible and two-dimensional 
(axisymmetric, finite stars). In Papers~I and~II, the Newtonian MHD equations 
were evolved while here we evolved the relativistic 
MHD equations with the assumption that the initial poloidal fields is 
large compared 
to rotational energy, but weak compared to the gravitational 
binding energy. In Papers~I and~II, it was found that magnetic 
braking generates coherent toroidal magnetic fields and laminar velocity flows. 
In the models considered here, we found that the field breaks down because of 
the phase mixing both in Newtonian and relativistic regimes. Our analysis 
indicates that the phase mixing is 
a result of the geometry of the stars and the poloidal field lines: 
phase mixing is absent in the infinite cylinders of Papers~I and~II 
simply because the systems are highly symmetric. However, even in finite 
stars, if viscosity is sufficiently strong, the toroidal fields may be 
damped before phase mixing becomes appreciable.

Differential rotation is damped by the presence of poloidal magnetic 
fields and viscosity. Magnetic braking causes the toroidal 
fields and angular velocities to undergo oscillations along each poloidal 
field line. In the regime explored in this paper, the oscillations along a 
given field line are independent of those along other lines. 
This phase mixing effect is likely to stir 
up turbulent-like motion. When viscosity is present, the star ultimately is driven 
toward uniform rotation. When the compactness of the star, $M/R$, is 
not very large, the effect of magnetic braking in a relativistic star is 
similar to the Newtonian 
case, but the timescale of the braking process is increased roughly by 
a factor $(1-2M/R)^2$. 
When $M/R$ approaches the Buchdahl limit $4/9$, 
the braking process is strongly affected by the spacetime curvature. Strong 
toroidal fields pile up at small $r$ along each field line, since the proper 
time elapses more slowly than at larger $r$ for the same amount of 
coordinate time. The whole process of 
magnetic braking can take much longer in this case than in the Newtonian situation. 

The assumptions necessary to bring the 2+1 MHD equations into
a set of 1+1 equations are that the system is axisymmetric; viscosity, 
magnetic diffusion and 
meridional currents can be neglected; and both magnetic and rotational
energies are much smaller than the gravitational binding energy.
The condition $E_{\rm mag} \gg E_{\rm rot}$ is a sufficient but 
may not be a necessary 
condition for the meridional currents to remain small in tens of Alfv\'en 
times. It is possible that the meridional currents can  
still be small compared to the rotational velocity even if this condition is 
not satisfied. If this is the case, our analysis here will still be 
valid. 
It is possible that most young neutron stars or merged binary neutron star 
remnants do not have strong poloidal
fields and the opposite limit $E_{\rm rot} \gg E_{\rm mag}$ may be more relevant. 
More detailed MHD calculations are required to study the general case.

The microscopic viscous timescale is
of order $t_{\nu} = R^2/8\nu \approx 10^9~{\rm s}$ for a typical neutron 
star, where $\nu = 347 \rho^{9/4} T^{-2}~{\rm cm}^2~{\rm s}^{-1}$ 
and $T$ is temperature~\cite{cutler87}. However, 
turbulent viscosity can be much bigger and might drive the star 
toward uniform rotation on much shorter timescale. The Alfv\'en 
timescale in a neutron star with magnetic field $B = 10^{12}~{\rm G}$ 
is tens of seconds (Paper~I). Hence, the effect of 
magnetic field is much more important than microscopic viscosity in 
neutron stars. Magnetic braking may have important consequences for 
gravitational wave signals and 
gamma-ray bursts, as discussed in the Introduction.

A significant amount of meridional current may also be 
generated as a result of the phase mixing. 
Our calculations do not take into 
account the later build-up of meridional currents, which may induce convective 
instabilities~\cite{pons99}, driving the seed magnetic field to high 
values greatly exceeding $10^{12}~{\rm G}$~\cite{duncan92}. The meridional 
currents may also induce other possible MHD instabilities, which may 
contribute to the redistribution of angular momentum~\cite{balbus98,spruit99}. 
Our goal here is to show that even in the simplest case, 
magnetic braking can also induce irregular, turbulent-like behavior.

More realistic evolutionary calculations of magnetic braking in neutron stars 
should clarify some of the above issues. One computational subtlety is that 
the Alfv\'en timescale is usually much longer than the dynamical timescale 
of the star. In this regime it may prove too taxing to a relativistic MHD 
code to evolve a differentially rotating star for the required length of 
time for magnetic braking to take effect. One possibility is to treat part 
of the evolution in the quasistatic approximation, as in a typical stellar 
evolution code, up to the moment that stable equilibrium can no longer be 
sustained. One other possibility is to artificially amplify
the magnetic field so that the effect of magnetic braking will show up
in a computationally managable timescale, and then scale the results 
for smaller ratios. Still 
another approach is to use implicit differencing scheme to avoid the
Courant constraint on the evolution timestep.
However, our calculations suggest that small-scale irregular angular 
velocity flows and meridional currents are likely to grow during the 
process of magnetic braking. The ability of a numerical code (finite 
difference or spectral) to resolve this behavior on ever decreasing 
scales is an important challenge that will require further analysis.
We hope to tackle the more general problems in the near future by means 
of numerical simulations in full general relativity.

We thank Charles Gammie and Jon McKinney for useful discussions. 
This work was supported 
in part by NSF Grants PHY-0090310 and PHY-0205155 and NASA Grant 
\mbox{NAG~5-10781} at UIUC.

\appendix

\section{Validity of Ignoring $\ve{v^r}$ and $\ve{v^{\theta}}$}
\label{app:v}

The main effect of magnetic braking is to wind up the frozen-in magnetic 
field and generate toroidal fields, which back-react on the angular velocity, 
$v^{\phi}=\Omega$. 
The change in angular velocity ultimately drives a meridional current.  
In this appendix, 
we estimate the timescale for $v^{\hat{r}}$ and $v^{\hat{\theta}}$ to become 
comparable to $v^{\hat{\phi}}$. Here we only study the simplest case: 
a differentially rotating, incompressible Newtonian star, with an initial $l=1$ 
poloidal field $\ve{B}(0)=B_0 \ve{e}_z$, and zero viscosity.

We start from the Newtonian MHD equations~(\ref{eq:beq1})--(\ref{eq:beq4}). 
We neglect viscosity in this appendix.
We find it convenient to work in cylindrical coordinates ($\varpi,\phi,z$). 
The meridional 
components of the velocity are $v^{\varpi}=\sin \theta v^{\hat{r}} + \cos \theta 
v^{\hat{\theta}}$, and $v^z=\cos \theta v^{\hat{r}} - \sin \theta v^{\hat{\theta}}$.
The MHD equations become 
\beqn
  \partial_t B^{\phi} &=& B_0 \partial_z \Omega + [ (B^z-B_0) \partial_z \Omega 
 + B^{\varpi} \partial_{\varpi} \Omega \cr  
 & &  - v^j \partial_j B^{\phi} ] \ ,
\label{eq:fulleq1} \\
  \partial_t \Omega &=& \frac{B_0}{4\pi\rho}\partial_z B^{\phi} + 
\left[ \frac{1}{4\pi\rho} \left( B^{\varpi} \partial_{\varpi} B^{\phi} + 
\frac{2}{\varpi}B^{\varpi} B^{\phi} \right) \right. \cr 
 & & \left. -v^j \partial_j \Omega  
 - \frac{2}{\varpi} v^{\varpi} \Omega \right] \ , \label{eq:fulleq2} \\
  \partial_t B^i &=& B^j \partial_j v^i - v^j \partial_j B^i \ , \\
  \partial_t v^{\varpi} &=& -v^j \partial_j v^{\varpi} +\varpi \Omega^2 -
\frac{1}{\rho} \partial_{\varpi} P - \partial_{\varpi} \Phi \cr
 & &  -\frac{1}{8\pi \rho} \partial_{\varpi} B^2 
+ \frac{1}{4\pi \rho} [ B^j \partial_j B^{\varpi} - \varpi (B^{\phi})^2]
\label{eq:fulleq4}  \ ,\\
 \partial_t v^z &=& -v^j \partial_j v^z - \frac{1}{\rho} \partial_z P 
- \partial_z \Phi -\frac{1}{8\pi \rho} \partial_z B^2 \cr 
& &  + \frac{1}{4\pi \rho} B^j \partial_j B^z \ , 
\label{eq:fulleq5} \\
  \partial_t B^{\varpi} &=& v^j \partial_j B^{\varpi} - B^j \partial_j v^{\varpi} \ , \\
  \partial_t B^z &=& v^j \partial_j B^z- B^j \partial_j v^z \ ,
\label{eq:fulleqn}
\eeqn
where $i$ and $j$ denotes $\varpi$ and $z$, and the summation convention is assumed. 
At $t=0$, 
we assume $v^{\varpi}(0;\varpi,z)=v^z(0;\varpi,z)=0$, 
$v^{\phi}(0;\varpi,z)=\Omega(0;\varpi,z)$, $\ve{B}(0;\varpi,z)=B_0 \ve{e}_z$, and 
the star is in hydrostatic equilibrium:
\beqn
  \varpi \Omega^2(0;\varpi,z)-\frac{1}{\rho} \partial_{\varpi} 
P(0;\varpi,z) - \partial_{\varpi} \Phi(0;\varpi,z) &=& 0 \ , \ \ \ \ \ \ \ \ 
\label{eq:hydroeq1} \\
  -\frac{1}{\rho} \partial_z P(0;\varpi,z) - \partial_z \Phi(0;\varpi,z) &=& 0  \ ,
\label{eq:hydroeq2}
\eeqn
where we have used Eqs.~(\ref{eq:fulleq4}) and~(\ref{eq:fulleq5}). After grouping 
these equilibrium terms, we see that the only remaining nonvanishing term at $t=0$ in 
Eqs.~(\ref{eq:fulleq1})--(\ref{eq:fulleqn}) is $B_0 \partial_z \Omega$, the 
first term on the right side of Eq.~(\ref{eq:fulleq1}). This simply tells us 
the obvious fact
that when there is differential rotation along the initial field lines (which is 
in the $z$-direction for this initial field), a toroidal field $B^{\phi}$ will 
be generated by a term linear in $\Omega$. It follows from Eq.~(\ref{eq:fulleq2}) 
that the toroidal field will then modify $\Omega$. Given our assumptions of slow 
rotation and weak magnetic field,
all the other terms are of higher order, at least at 
small $t$. This is the reason we neglect them in this paper. However, it is possible
that $v^{\varpi}$ and $v^z$ will become important at late times.

Here we crudely estimate the timescale in which $v^{\varpi}$ and $v^z$ 
are driven to a magnitude comparable to $v^{\hat{\phi}}$ by the change of 
$\Omega$ induced by magnetic braking. We focus on Eq.~(\ref{eq:fulleq4}).
We assume that the pressure $P$ and gravitational 
potential $\Phi$ in our nearly spherical configuration do not change and 
set $v^{\varpi}=v^z=0$, $B^{\varpi}=0$, 
and $B^z=B_0$ on the right side of Eq.~(\ref{eq:fulleq4}). Using Eq.~(\ref{eq:hydroeq1}),
Eq.~(\ref{eq:fulleq4}) becomes 
\beq
  \partial_t v^{\varpi} = \varpi [\Omega^2(t)-\Omega^2(0)] 
 - \frac{[B^{\hat{\phi}}(t)]^2}{4\pi \rho \varpi} \ .
\label{eq:oversim}
\eeq
We then substitute for
$\Omega$ and $B^{\phi}$ the analytic solutions given by 
Eqs.~(\ref{eq:analBphi}) and~(\ref{eq:analOmega}). For simplicity, 
we assume that only one mode is present, and write 
\beqn
  \Omega(t) &=& \Omega(0) + \epsilon 
\left( 1- \frac{\varpi^2}{R^2} \right)
\cos k_n z \, (1-\cos \omega_n t) \ , \label{eq:Omega1} \ \ \ \ \ \ \ \  \\
  B_1^{\hat{\phi}}(t) &=& B_0 \left( \frac{\varpi \epsilon}{v_A}\right) 
\left( 1- \frac{\varpi^2}{R^2} \right) \sin k_n z \, \sin \omega_n t \ , 
\label{eq:B1phi}
\eeqn
where $\epsilon$ is a constant parameter of dimension $1/{\rm time}$, measuring the
degree of differential rotation along the initial magnetic field lines. We 
assume that the star may have a high degree of differential rotation so that 
$\epsilon$ can be of the same order as $\Omega(0)$. The factor $(1-\varpi^2/R^2)$ 
is inserted to make both $\Omega$ and $B^{\hat{\phi}}$ regular at $\varpi=R$.
Equation~(\ref{eq:oversim}) can now be solved 
analytically. The growth of $v^{\varpi}$ is dominated by the terms that 
go linearly in time:
\beqn
  v^{\varpi} &\approx& t \varpi \epsilon^2 \left(1-\frac{\varpi^2}{R^2}\right) 
\left[ \frac{2\Omega(0)}{\epsilon} \cos k_n z \right. \cr 
& & \left. + 2\left(1-\frac{\varpi^2}{R^2}\right)
\cos^2 k_n z -\frac{1}{2} \left(1-\frac{\varpi^2}{R^2}\right) \right] \ . 
\ \ \ \ \ \ \ \ 
\eeqn
Hence $v^{\varpi}$ will be comparable to $v^{\hat{\phi}} \sim \Omega(0)\varpi$ when 
$t\ga t_c \sim \Omega(0)/\epsilon^2 > 1/\Omega(0)$. Hence $t_c \ga P_{\rm rot}$, 
where $P_{\rm rot}$ is a rotation period. 
It follows from the incompressibility condition $\ve{\nabla} \cdot \ve{v}=0$ 
that $v^z$ is in general of the same order as $v^{\varpi}$. Therefore, 
$v^z$ will also develop on the timescale $t_c$. On the other hand, the timescale 
for magnetic braking to affect $v^{\hat{\phi}}$ and generate $B^{\hat{\phi}}$ 
is $t_A$, the Alfv\'en timescale.
Although our calculation is 
done on a Newtonian model with an $l=1$ initial field, we expect that the 
result is similar in general relativity and with other initial magnetic field 
configurations.

Our estimate is based on dropping all the terms on the right side of 
Eq.~(\ref{eq:fulleq4}) which are of higher order at early time. This crude 
calculation therefore does not take into account the possibility of the growth 
of other instabilities, like
the magnetorotational instability (MRI). However, the growth time of 
this instability is also of order $P_{\rm rot}$ (see e.g., Ref.~\cite{balbus98}, 
Section~IV). 
Hence we conclude that our numerical results are valid as long as the Alfv\'en 
timescale $t_A \ll P_{\rm rot}$, or equivalently $E_{\rm rot} \ll E_{\rm mag}$.

Although our analysis in this paper is restricted to the early phases
of the evolution $P_{\rm rot} \gg t \ga t_A$ in the slow rotation, weak 
magnetic field limit,
we are able to determine the nonlinear evolution of the
angular velocity and toroidal magnetic field profiles
during this period for a fully relativistic, differentially rotating star.

\section{Transformation between $\ve{\hat{u}}$, $\ve{\hat{v}}$, 
$\ve{r}$ and $\ve{\theta}$ for $\ve{l=2}$ Field}
\label{app:uv}

To solve Eqs.~(\ref{eq:dtBl2}) and~(\ref{eq:dtWl2}), we need to know the 
transformation between $r$, $\theta$ and $\hat{u}$, $\hat{v}$. In this appendix, 
we show the required transformation.

It is convenient to introduce the cylindrical variables 
\beq
  \hat{\varpi} = \hat{r} \sin \theta  \ \ , \ \ 
  \hat{z} = \hat{r} \cos \theta \ ,
\label{eq:pm-z}
\eeq 
where $\hat{r}=r/R$. We also define an auxiliary variable 
\beq
  \hat{u}_* = u/R^2 = -\hat{r}^2 (3 \cos^2 \theta - 1) \ .
\eeq
The relations between $\hat{\varpi}$, $\hat{z}$, and $\hat{u}_*$, $\hat{v}$ 
are 
\beqn
  \hat{u}_* &=& \hat{\varpi}^2-2\hat{z}^2 \ , \label{eq:ustar} \\
  \hat{v} &=& \frac{3\sqrt{3}}{2} \hat{\varpi}^2 \hat{z} \ .
\label{eq:vhat}
\eeqn
For a given value of $\hat{v}$, we denote by $\hat{\varpi}_1$ and $\hat{\varpi}_2$ 
(with $\hat{\varpi}_2 > \hat{\varpi}_1$)
the two points at which the $\hat{v}=$constant line intercepts the surface of 
the star $\hat{r}=1$. Substituting $\hat{z}=\sqrt{1-\hat{\varpi}^2}$ in 
Eq.~(\ref{eq:vhat}), we obtain the equation for 
$\hat{\varpi}_1$ and $\hat{\varpi}_2$: 
\beq	
  \hat{v}=\frac{3\sqrt{3}}{2} \hat{\varpi}^2 \sqrt{1-\hat{\varpi}^2} \ .
\eeq
The equation can be turned into a cubic equation and the solution is given by 
\beqn
  \hat{\varpi}_1 &=& \sqrt{\frac{1}{3} \left[ 1 + 2 \cos \left( 
\psi + \frac{4}{3}\pi \right) \right]} \ , \label{eq:pomega1} \\ 
  \hat{\varpi}_2 &=& \sqrt{ \frac{1}{3} (1 + 2 \cos \psi)} \ , \label{eq:pomega2} \\
  \psi &=& \frac{2}{3} \sin^{-1} \hat{v} \ .
\eeqn
It follows from Eqs.~(\ref{eq:ustar}) and~(\ref{eq:vhat}) that the corresponding 
$\hat{u}_{1*}$ and $\hat{u}_{2*}$ are given by 
\beq
  \hat{u}_{j*} = \hat{\varpi}_j^2 - \frac{8 \hat{v}^2}{27 \hat{\varpi}_j^4} \ \ , \  \ 
(j=1, \ 2) \ . \label{eq:u1u2star}
\eeq

The transformation from $(r,\theta)$ to $(\hat{u},\hat{v})$ 
can be performed by the following procedure. We first compute $\hat{u}_*$ and 
$\hat{v}$ from Eqs.~(\ref{eq:pm-z})--(\ref{eq:vhat}). We then compute 
$\hat{u}_{1*}$ and $\hat{u}_{2*}$ from Eqs.~(\ref{eq:pomega1})--(\ref{eq:u1u2star}). 
Finally, we compute $\hat{u}$ by the formula [see Eq.~(\ref{eq:uhat})]
\beq
  \hat{u} = \frac{\hat{u}_*-\hat{u}_{1*}}{\hat{u}_{2*}-\hat{u}_{1*}} \ .
\label{eq:hatu2}
\eeq

To transform from $(\hat{u},\hat{v})$ to $(r,\theta)$, we first compute 
$\hat{u}_{1*}$ and $\hat{u}_{2*}$ from Eqs.~(\ref{eq:pomega1})--(\ref{eq:u1u2star}). 
Next we calculate $\hat{u}_*$ from Eq.~(\ref{eq:hatu2}). We then solve 
Eqs.~(\ref{eq:ustar}) and~(\ref{eq:vhat}) to obtain $\hat{\varpi}$ and 
$\hat{z}$. The two equations can be combined into a cubic equation, which 
is solved by the standard cubic equation formula. Finally, the values of $r$ 
and $\theta$ is obtained from Eq.~(\ref{eq:pm-z}).

\section{Finite differencing scheme}
\label{app:FD}

In this appendix, we describe in details our finite differencing scheme 
for the $l=1$ MHD equations~(\ref{eq:dtBl1}) and~(\ref{eq:dtWl1}) in 
the presence of viscosity. The other cases are treated in a similar fashion. 

We define
\beq
  X^n_{i,j} = X(\hat{t}^n;\hat{u}_i,\hat{\varpi}_j) \ ,
\eeq
where $X$ is any variables and 
\beqn
  \hat{t}^n &=& n \Delta \hat{t} \ \ , \ \ \hat{u}_i = i \Delta \hat{u} \ \ , \ \ 
\hat{\varpi}_j = j \Delta \hat{\varpi} \ \ .
\eeqn
We use a uniform spatial grid ($\Delta \hat{u}=$ constant, $\Delta \hat{\varpi}=$
constant) and a uniform time step ($\Delta \hat{t}=$constant) in our calculations. 
We store $\hat{B}$ at grid points $(i,j)$ and $\hat{\Omega}$ at $(i+1/2,j)$. 
We compute the right side of Eqs.~(\ref{eq:dtBl1}) and~(\ref{eq:dtWl1}) 
by the following finite difference approximation: 
\beqn 
  \partial_{\hat{t}} \hat{B} (t^n;\hat{u}_i,\hat{\varpi}_j) &\approx & 
f_{B;i,j}(\hat{\Omega}^n) \ , \label{eq:FDdB} \\
  \partial_{\hat{t}} \hat{\Omega} (t^n;\hat{u}_{i+1/2},\hat{\varpi}_j) &\approx &
f_{\Omega;i+1/2,j}(\hat{B}^n) \cr 
& & + f_{v;i+1/2,j}(\hat{\Omega}^n)  \ ,
\label{eq:FDdW}
\eeqn
where 
\beqn
  f_{B;i,j}(\hat{\Omega}^n) &=& \frac{\alpha_{i,j} \lambda_{i,j}}
{\sqrt{1-\hat{\varpi}_j^2}} \left( \frac{\hat{\Omega}^n_{i+1/2,j}
-\hat{\Omega}^n_{i-1/2,j}}{\Delta \hat{u}} \right)  \\
  f_{\Omega;i+1/2,j}(\hat{B}^n) &=& \frac{\alpha_{i+1/2,j}^2 \lambda_{i+1/2,j}}
{\alpha_R \sqrt{1-\hat{\varpi}_j^2}} \left( \frac{\hat{B}^n_{i+1,j}
-\hat{B}^n_{i,j}}{\Delta \hat{u}} \right) \ \ \ \ \ \ 
\eeqn
We use the expression in Eq.~(\ref{eq:visterm}) to compute the finite 
difference term, $f_{v;i+1/2,j}(\hat{\Omega}^n)$, for 
$[\partial_{\hat{t}} \hat{\Omega}(t^n;\hat{u}_{i+1/2},\hat{\varpi}_j)]_{\rm vis}$. 
Specifically, we first calculate 
$\partial_{\hat{u}} \hat{\Omega}$, $\partial_{\hat{u}}^2 \hat{\Omega}$,
$\partial_{\hat{\varpi}} \hat{\Omega}$, $\partial_{\hat{\varpi}}^2
\hat{\Omega}$ and $\partial_{\hat{\varpi}} \partial_{\hat{u}}
\hat{\Omega}$ by standard central differencing, and then
compute $\partial_{\hat{r}} \hat{\Omega}$, $\partial_{\hat{r}}^2
\hat{\Omega}$, $\partial_{\theta} \hat{\Omega}$ and $\partial_{\theta}^2
\hat{\Omega}$ by the transformation formulae
\beqn
 \partial_{\hat{r}} \hat{\Omega} &=& \frac{\hat{\varpi}}{\hat{r}}
\partial_{\hat{\varpi}} \hat{\Omega} + \frac{\hat{u}}{\hat{r}
(1-\hat{\varpi}^2)} \partial_{\hat{u}} \hat{\Omega} \label{eq:dromega} \\
 \partial_{\theta} \hat{\Omega} &=& \hat{u} \sqrt{1-\hat{\varpi}^2}
\partial_{\hat{\varpi}} \hat{\Omega} - \frac{\hat{\varpi}}
{\sqrt{1-\hat{\varpi}^2}} (1-\hat{u}^2) \partial_{\hat{u}} \hat{\Omega} \\
 \partial_{\hat{r}}^2 \hat{\Omega} &=& \frac{\hat{\varpi}^2}{\hat{r}^2}
\partial_{\hat{\varpi}}^2 \hat{\Omega} + \frac{\hat{u}^2}{\hat{r}^2
(1-\hat{\varpi}^2)^2} \partial_{\hat{u}}^2 \hat{\Omega}
+ \frac{2 \hat{\varpi} \hat{u} }{\hat{r}^2 (1-\hat{\varpi}^2)}
\partial_{\hat{\varpi}} \partial_{\hat{u}} \hat{\Omega}  \cr
 & & + \frac{3 \hat{\varpi}^2 \hat{u} }{\hat{r}^2 (1-\hat{\varpi}^2)^2}
\partial_{\hat{u}} \hat{\Omega} \\
  \partial_{\theta}^2 \hat{\Omega} &=& \hat{u}^2 (1-\hat{\varpi}^2)^2
\partial_{\hat{\varpi}}^2 \hat{\Omega} + \frac{\hat{\varpi}^2}
{1-\hat{\varpi}^2}(1-\hat{u}^2)^2 \partial_{\hat{u}}^2 \hat{\Omega} \cr & & \cr
 & & - 2 \hat{\varpi} \hat{u}(1-\hat{u}^2) \partial_{\hat{\varpi}}
\partial_{\hat{u}} \hat{\Omega} 
  -\hat{\varpi} \partial_{\hat{\varpi}} \hat{\Omega} \cr & &  \cr & & 
  - \hat{u} (1-\hat{u}^2) \frac{1+2 \hat{\varpi}^2}{1-\hat{\varpi}^2}
\partial_{\hat{u}} \hat{\Omega} \ . 
\label{eq:dth2omega}
\eeqn
Finally, we compute $f_{v;i+1/2,j}(\hat{\Omega}^n)$ using 
Eq..~(\ref{eq:visterm}).

We use an iterated Crank-Nicholson~\cite{teukolsky00} scheme to evolve 
Eqs.~(\ref{eq:FDdB}) and~(\ref{eq:FDdW}).
To be specific, we first
predict $\hat{B}$ and $\hat{\Omega}$ at the next time step by  
\beqn
  { ^{(1)}}\hat{B}^{n+1}_{i,j} &=& \hat{B}^n_{i,j} + \Delta t 
f_{B;i,j}(\hat{\Omega}^n) \ , \\
  { ^{(1)}}\hat{\Omega}^{n+1}_{i+1/2,j} &=& \hat{\Omega}^n_{i+1/2,j} 
+ \Delta t [f_{B;i+1/2,j}(\hat{B}^n) \cr 
& &  + f_{v;i+1/2,j}(\hat{\Omega}^n)] \ .
\eeqn
Next we estimate $\hat{B}$ and $\hat{\Omega}$ at the next half time 
step by 
\beqn
  { ^{(1)}}\hat{B}^{n+1/2}_{i,j} &=& \frac{1}{2} \left[ \hat{B}^n_{i,j}+
{ ^{(1)}}\hat{B}^{n+1}_{i,j} \right] \ , \\
  { ^{(1)}}\hat{\Omega}^{n+1/2}_{i+1/2,j} &=& \frac{1}{2} \left[ 
\hat{\Omega}^n_{i+1/2,j} + { ^{(1)}}\hat{\Omega}^{n+1}_{i+1/2,j} \right ] \ .
\eeqn
We then use these variables to obtain a more accurate estimate of $\hat{B}$ and 
$\hat{\Omega}$ at the next time step: 
\beqn
  { ^{(2)}}\hat{B}^{n+1}_{i,j} &=& \hat{B}^n_{i,j} + \Delta t
f_{B;i,j}({ ^{(1)}}\hat{\Omega}^{n+1/2}) \ , \\
  { ^{(2)}}\hat{\Omega}^{n+1}_{i+1/2,j} &=& \hat{\Omega}^n_{i+1/2,j}
+ \Delta t [f_{B;i+1/2,j}({ ^{(1)}}\hat{B}^{n+1/2}) + \cr & & 
f_{v;i+1/2,j}({ ^{(1)}}\hat{\Omega}^{n+1/2})] \ .
\eeqn
To ensure the stability of this finite differencing scheme, we have to do 
this ``corrector'' step one more time~\cite{teukolsky00}. Hence we compute 
\beqn
  { ^{(2)}}\hat{B}^{n+1/2}_{i,j} &=& \frac{1}{2} \left[ \hat{B}^n_{i,j}+
{ ^{(2)}}\hat{B}^{n+1}_{i,j} \right] \ , \\
  { ^{(2)}}\hat{\Omega}^{n+1/2}_{i+1/2,j} &=& \frac{1}{2} \left[
\hat{\Omega}^n_{i+1/2,j} + { ^{(2)}}\hat{\Omega}^{n+1}_{i+1/2,j} \right ] \ ,
\eeqn
and use them to obtain the final values of $\hat{B}$ and $\hat{\Omega}$ at the 
next time step: 
\beqn
  \hat{B}^{n+1}_{i,j} &=& \hat{B}^n_{i,j} + \Delta t
f_{B;i,j}({ ^{(2)}}\hat{\Omega}^{n+1/2}) \ , \\
  \hat{\Omega}^{n+1}_{i+1/2,j} &=& \hat{\Omega}^n_{i+1/2,j}
+ \Delta t [f_{B;i+1/2,j}({ ^{(2)}}\hat{B}^{n+1/2}) \cr 
& & + f_{v;i+1/2,j}({ ^{(2)}}\hat{\Omega}^{n+1/2})] \ .
\eeqn
The iterated Crank-Nicholson scheme is second order accurate in space and time.

\end{document}